# Time Varying Risk in U.S. Housing Sector and Real Estate Investment Trusts Equity Return

Masud Alam[1]

Abstract:

This study examines how housing sector volatilities affect real estate investment trust (REIT) equity return in the United States. I argue that unexpected changes in housing variables can be a source of aggregate housing risk, and the first principal component extracted from the volatilities of U.S. housing variables can predict the expected REIT equity returns. I propose and construct a factor-based housing risk index as an additional factor in asset price models that uses the time-varying conditional volatility of housing variables within the U.S. housing sector. The findings show that the proposed housing risk index is economically and theoretically consistent with the risk-return relationship of the conditional Intertemporal Capital Asset Pricing Model (ICAPM) of Merton (1973), which predicts an average maximum of 5.6 percent of risk premium in REIT equity return. In subsample analyses, the positive relationship is not affected by sample periods' choice but shows higher housing risk beta values for the 2009-18 sample period. The relationship remains significant after controlling for VIX, Fama-French three factors and a broad set of macroeconomic and financial variables. Moreover, the proposed housing beta also accurately forecasts U.S. macroeconomic and financial conditions.

---

[1] Northern Illinois University, IL, USA, & Shahjalal University of Science and Technology, Bangladesh. Email: masud.sust@gmail.com

# Time Varying Risk in U.S. Housing Sector and Real Estate Investment Trusts Equity Return

Masud Alam[1]

## 1. Introduction

Real Estate Investment Trusts (REITs) are tradeable income-generating real estate equity investment vehicles related to the housing sector and are listed on major exchanges with more than $3 trillion in gross assets and more than $1 trillion equity market capitalization across a range of U.S. real estate sectors. In 2019, approximately 87 million U.S. investors invested in REITs stocks through their 401(k)[2] and these portfolio investments are directly associated with more than 500,000 properties. Therefore, understanding what variables drive the fluctuation of real estate investment trusts (REIT) equity return is of paramount importance for REIT investors aiming to make optimal investment decisions.

Existing studies mainly focus on monetary and fiscal policy shocks, firm-specific idiosyncratic risk, financial, and standard equity risk variables in explaining real estate stock and REIT returns. For example, Bond and Xue (2017) consider firm-specific investment and profitability factors to predict the expected REIT returns, Liow, Ibrahim, and Huang (2006) consider conditional volatilities of six macroeconomic variables and their expected risk premia on property stock returns, and McCue and Kling (1994) examine the response of REIT returns to shocks related to four macroeconomic variables. While the dynamics in conditional correlations of REIT returns and the volatility of macroeconomic predictors is well examined in the empirical literature, studies on the role of predictors directly related to the U.S. housing market is limited.

Recent theoretical studies show that the volatility associated with the housing sector is not adequately identified as a distinct risk factor in the capital asset pricing model (CAPM) or in Fama-French three factors model (DeLisle, Price, & Sirmans, 2013; Ooi, Wang & Webb, 2009; Sun, Titman, &Twite, 2015) when applied to REIT return estimation. Research by Morck, Yeung, and Yu (2000) and Roll (1988) also observe that the factors related to economic and market fundamentals can only explain a small portion of the asset return volatility and suggest sector-specific factors can better explain the remaining unexplained variance of asset returns. Given the lack of literature on housing volatilities and the theoretical call for more sector-specific factors in the asset pricing model, this study focuses on identifying relevant factors within the U.S. housing sector and examines if the conditional covariance of volatilities in the housing sector with those factors and REIT returns are significantly correlated.

This study proposes a housing risk index as an additional measure of the volatilities within the U.S. housing sector. A two-step methodology is used to estimate the conditional volatilities of these ten housing variables. In the first step, common factors are extracted using a dynamic factor model (Alam, 2021; Bernanke, Boivin, & Eliasz, 2005) from an extensive macroeconomic data set, made up of 76 U.S. monthly macroeconomic series. Next, we estimate ten separate bivariate Threshold-Generalized Autoregressive Conditional Heteroscedastic (T-GARCH) models (Glosten, Jagannathan, & Runkle, 1993), which include the first common factor[3] and one of the ten housing variables to account for asymmetric responses of the volatility of housing variables. This yields a conditional volatility series for the ten housing variables from which a risk index is constructed by estimating the first principal component of the conditional volatility series of the ten housing variables.

The conditional volatility of the ten housing variables is chosen after being identified in the previous literature as fundamental variables of the U.S. housing market. For instance, Fratantoni and Schuh (2003) use regional housing starts and mortgage rate, Davis and Heathcote (2005) examine mortgage markets and Case-Shiller house prices, Saks (2008) uses newly constructed housing units and the number of houses sold, and Mian and Sufi (2009) examine the role of subprime mortgage credit to estimate the sources of aggregate housing market volatility.

The results of the univariate regressions support the positive risk-return relationship of the intertemporal capital asset pricing model (ICAPM) (Merton, 1973) when the constructed housing risk index is included as a state variable. The estimated housing risk betas are economically and theoretically consistent across the cross-section of regressions of ICAPM models and predict, on average, a maximum 5.6 % risk premium for REIT equity returns. This risk-return relationship remains stable and significant when regression models include controls for a set of macroeconomic and financial variables. Consistent with Chiang and Li (2012), the quantile risk betas show a higher positive risk-return relation when the distribution of returns is above the 10% quantile. After confirming the housing risk index's economic and statistical significance across univariate and multivariate cross-section regressions, this study examines the predictive power of the housing risk index for the U.S. business cycles fluctuations and finds that the proposed housing risk index accurately forecasts the U.S. macroeconomic and financial conditions and significantly predicts regional macroeconomic fluctuations.

---

[3] The first factor explains around 73 percent of total variation of data. The one-factor T-GARCH model is also applied to estimate the return and variance dynamics of German stocks by Kaiser (1996) that is also similar to Diebold and Nerlove (1989) and Engle, Ng, and Rothschild, (1990).

These findings are robust across models after controlling for a set of standard macroeconomic and financial variables from Welch and Goyal (2008). The model further controls for the Fama-French three factors and find economically and statistically significant evidence of housing risk premia in all regression specifications. The maximum and minimum housing risk premia across all multivariate OLS models are an average of 3.7 percent and 1.9 percent, respectively. The housing risk premium remains almost stable in quantile regression with a maximum of 6.7 percent for the 95[th] percentile and 5.6 percent for the 75[th] percentile of the return distribution.[4]

In addition to in-sample predictive analysis, the analysis includes an out-of-sample and sub-sample forecasting analysis. Over the out-of-sample periods, the estimated coefficient of housing risk beta remains positive and consistent with the in-sample estimates and deliver lower out-of-sample predictive errors for the Fama-French three-factor model. The sub-sample analysis also supports the robust relationship between the housing risk index and the expected REIT returns. The analysis uses April, 2009 as a breakpoint and shows that the positive relationship remains unaffected by sample periods' choice but shows higher values of housing risk beta for the May, 2009 – December, 2018 sample period

---

[4] Springer and Cheng (2006) estimate firm-specific risk premium ranging from 2% to 9% for individual REITs, Edelstein and Magin (2013) show U.S. REIT equity risk premium is likely in between 4.3 and 6.3 percent. The historical REITs risk premium in excess of the 10-year treasury bond is approximately 6% for 1972-2015 period (Van Nieuwerburgh, 2019).

## 2. Related literature

This study complements and contributes to three parts of the asset price literature. First, this paper is related to the literature concerning the construction of the new risk factors by combining a high-dimensional factor model with the GARCH model (Bali, Brown, & Caglayan, 2014; Ghysels, Plazzi, Valkanov, & Torous, 2013; Plazzi, Torous, & Valkanov, 2010) and accounting for asymmetric responses of the volatility of housing variables, as in Kwan, Li, and Ng, (2005) and Liow, Zhou, and Ye, (2015). In this context, this study contributes to a large strand of REIT literature that has explored the empirical applications of principal component and latent factor approach in arbitrage pricing theory (Lizieri, Satchell, & Zhang, 2007; Mühlhofer, 2013; and Kozak, Nagel, & Santosh, 2018). Second, the factor-based bi-variate T-GARCH approach employed in this study contributes to the improvement of existing multivariate GARCH estimates of the sector-specific risk factors and examining the role these risk factors in predicting REIT returns (Bali et al., 2014; Baillie and Myers, 1991; Chuang, Liu, Susmel, 2012; Devaney, 2001; Kaiser, 1996). Finally, the empirical examination of ICAPM and the Fama-French model in REIT returns lies at the confluence of broader asset price literature that examines the risk-return relationship and estimates expected risk premia of new risk factors (Chan, Hendershott, & Sanders, 1990; Devaney, 2001; Ewing and Payne, 2005; Liow et al., 2015; and Naranjo and Ryngaert, 1998).

The construction of housing risk factors of this study and the application of the factor-GARCH approach are closely related to Bali et al. (2014), Ghysels et al. (2013), Liow et al. (2006), and Plazzi et al. (2010). Bali et al. (2014) propose a broad index of a risk factors from eight macroeconomic variables and tests the proposed risk factor's predictive ability for hedge fund returns. They argue that unexpected changes in macroeconomic variables can be a source of aggregate risk, and the first principal component extracted from the volatilities of macroeconomic variables can predict the expected asset returns. Liow et al. (2006) apply principal component analysis with a GARCH (1,1) process to estimate the risk premia of macroeconomic risk factors in predicting the variation of REIT returns. They applied GARCH (1,1) approach to a set of estimated principal components of six macroeconomic variables and find that exposure to the six macroeconomic risk factors provides better predictions than the financial risk factors commonly used to explain REIT returns.

Evidence for predictability in REIT returns and commercial real estate returns is provided by Ghysels et al. (2013) and Plazzi et al. (2010) using principal components extracted from a set of macroeconomic risk factors. Lizieri et al. (2007) examine multifactor asset price models for the U.S. REIT returns, where macroeconomic risk factors are estimated using principal component and independent component analysis. They suggest that the multi-factor arbitrage pricing (APT) model with estimated macroeconomic risk factors outperforms other factor-based statistical models.[5] While a substantial volume of studies related to the construction of macroeconomic risk factors, no attention has been devoted to the U.S. housing sectors' risk factors. Furthermore, Leamer (2007) argues that housing is a business cycle considering its contribution to the U.S. GDP. This study fills this gap by considering variables from the U.S. housing sectors and estimates a housing risk index. The housing sector's risk is vital because the massive spike of volatility in housing sectors has identified as a main driving force in declining investment, prices, and employment in the U.S. economy during the great recession.

The bivariate T-GARCH approach employed in this study is related to econometric methodologies employed by Bali et al. (2014), Baillie and Myers (1991), Chuang et al. (2012). Devaney (2001), Kaiser (1996), Kwan, Li, and Ng (2005), and Liow et al. (2015). Liow et al. (2015) use the multivariate T-GARCH approach to model the macroeconomic risk variables in predicting property stock excess returns and Bali et al. (2014) use the same approach to estimate macroeconomic risk factors. Unlike Bali et al. (2014), who consider cross-correlation of eight correlated macroeconomic variables in the T-GARCH model, Kaiser (1996) employs a factor structure in the GARCH model with the assumption that the factors are orthogonal. Baillie and Myers (1991) focus on the joint distribution of the volatility of cash flow and the future prices of six commodities using the bivariate GARCH model, while Chuang et al. (2012) use the bivariate T-GARCH model to examine the covariance of ten Asian stock market's trading volume and return volatilities.

---

[5] Blundell and Ward (1987), Chen and Jordan (1993), Chen, Hsieh and Jordan (1997), and Titman and Warga (1986) also used the factor-analytic approach to estimate risk factors and examine the link between estimated risk factors and REIT returns in multifactor asset price models.

Among these examples, the methodology most similar to this study is that used by Bali et al. (2014), Baillie and Myers (1991), and Chuang et al. (2012). However, while Bali et al. (2014) estimate the model using a pre-specified eight macroeconomic variables, this study includes variables from U.S. housing sectors and a factor extracted from a large panel of macroeconomic variables. This study also estimates time-varying conditional mean volatility in contrast to the constant mean volatility estimate of Baillie and Myers (1991). Chuang et al. (2012) estimate the bivariate T-GARCH model without considering the asymmetric effects of macroeconomic innovations on the volatilities of asset returns, while this study incorporates time-varying effects of macroeconomic factors on REIT returns' volatilities. Additionally, the empirical procedure of this study differs from Devaney (2001) and Kwan et al. (2005) in selecting the lag of factors and housing variables.

This paper also contributes to the empirical literature examining multifactor asset pricing models linked with the volatility of macroeconomic variables. These include Merton's (1973) original intertemporal capital asset pricing models (ICAPM), a variant version of ICAPM by Cox, Ingersoll, and Ross (1985) and Rubinstein (1976), the conditional asset pricing model by Cochrane (1996), the consumption-based model by Campbell (1993), and the arbitrage pricing theory (APT) of Ross (1976). This literature group postulates a theoretical connection between expected stock returns and the default risk premium to one or more volatility measures of macroeconomic fundamentals. This connection motivates this study to hypothesize a positive relationship between the housing risk factors and excess REIT returns.

Chan et al. (1990) employ a multifactor asset pricing model to test how risk exposure of various factors influence the variation of U.S. stock, and REIT returns. The findings suggest that unexpected variations in macroeconomic factors significantly impact asset returns. The risk premium of macroeconomic factors in predicting expected REIT returns is also documented by Chen et al. (1997), Devaney (2001), and Naranjo and Ryngaert (1998) in various specifications of Fama-French asset pricing models. The consensus in these empirical studies is that the conditional variance of macroeconomic factors is priced into asset pricing models and predicts the higher expected future returns. The empirical validity of multifactor asset pricing models in the previous literature supports this study for assessing REIT returns exposures to housing risk factors.

On the predictability of U.S. REIT returns, Ewing and Payne (2005) and Liow et al. (2015) investigate whether equity REIT returns are related to the pre-specified macroeconomic factors. They find that those factors significantly predict the variation of REIT equity returns. In contrast to the pre-specified set of macroeconomic variables used in previous studies, this paper uses orthogonal macroeconomic factors from 76 macroeconomic variables and then estimate a housing risk factor by considering the joint dynamics of macroeconomic factors and the ten housing variables.

## 3. Econometric models and variables

### 3.1 Factor augmented bivariate T-GARCH model

The construction of the housing risk index for this study uses ten variables related to U.S. house prices, housing starts, and mortgage interest rates: (1) HOUST: total housing starts of new privately-owned housing units; (2) PERMIT: new private housing units authorized by building permits; (3) RHPI: S&P/Case-Shiller real U.S. national home price index; (4) HOUSTMW: housing starts in Midwest census region; (5) HOUSTNE: housing starts in northeast census region; (6) HOUSTS: housing starts in south census region; (7) HOUSTW: housing starts in west census region; (8) HOUSOLD: total number of houses sold in the U.S. Census regions; (9) MSACSR: the ratio of the house for sales to houses sold; and (10) MTSPRD: mortgage contract interest rate. The economics and finance literature (Belviso and Milani, 2005; Karolyi and Sanders, 1998; Leamer, 2007; McCue and Kling, 1994) consider these leading housing indicators, not only because they directly signal the dynamics of housing markets, but their volatilities also potentially affect the investment opportunities.

A bivariate Threshold Generalized Autoregressive Conditional Heteroscedasticity model with a time-varying parameter (T-GARCH) is used for these ten housing variables to estimate individual housing risk factors. In empirical implementation, I first estimate the conditional mean equation for an individual housing series ($X_{it}$) and the macroeconomic factor ($f_t$) using a standard vector auto-regression (VAR) model of order one:

$$\begin{bmatrix} X_{i,t} \\ f_t \end{bmatrix} = \begin{bmatrix} \beta_0^i \\ \beta_0^f \end{bmatrix} + \begin{bmatrix} \beta_1^i & \beta_2^i \\ \beta_1^f & \beta_2^f \end{bmatrix} \begin{bmatrix} X_{i,t-1} \\ f_{t-1} \end{bmatrix} + \begin{bmatrix} \varepsilon_{i,t} \\ \varepsilon_{f,t} \end{bmatrix} \tag{1}$$

where $X_{i,t}$ is one of the ten housing variables in $i$ at time $t$, $i = 1 \cdots 10$, and $f_i$ is a series of macroeconomic factor at time $t$, $\varepsilon_{it} \sim N(0, \sigma_{it}^2)$, $\varepsilon_{ft} \sim N(0, \sigma_{ft}^2)$, and $Cov\,(X_{it}, f_t) = \Sigma_{xf}$ The VAR (1) model in equation (1) accounts for cross-correlation of aggregate macroeconomic information through factors $(f)$ and the housing variables $(X)$. The factor is extracted from a large panel of macroeconomic time series data using principal component analysis. Let the information set $I_{t-1} = \{X_1^i, f_1 \ldots X_{t-1}^i, f_{t-1}\}$ that includes all the available past information of $X$ and $f$, the conditional volatility of $E(\varepsilon_{it}^2 \mid I_{t-1}) = \sigma_{it}^2$ and $E(\varepsilon_{ft}^2 \mid I_{t-1}) = \sigma_{ft}^2$, is estimated using the bi-variate T-GARCH models:

$$\sigma_{i,t}^2 = E[\varepsilon_{i,t}^2 \mid I_{t-1}] = \omega^i + \alpha^i\,\varepsilon_{i,t-1}^2 + \phi^i\,\sigma_{i,t-1}^2 + \gamma^i\,\varepsilon_{i,t-1}^2 D_{i,t-1} \tag{2}$$

$$\sigma_{f,t}^2 = E[\varepsilon_{f,t}^2 \mid I_{t-1}] = \omega^f + \alpha^f\,\varepsilon_{f,t-1}^2 + \phi^f\,\sigma_{f,t-1}^2 + \gamma^f\,\varepsilon_{f,t-1}^2 D_{f,t-1} \tag{3}$$

$$\sigma_{if,t} = E[\varepsilon_{i,t}\varepsilon_{f,t} \mid I_{t-1}] = \omega^{if} + \alpha^{if}\,\varepsilon_{i,t-1}\varepsilon_{f,t-1} + \phi^{if}\,\sigma_{if,t-1} + \gamma^{if}\,\varepsilon_{it-1}\,\varepsilon_{ft-1}\,D_{i,t-1}\,D_{f,t-1} \tag{4}$$

where $\omega > 0, \alpha \geq 0, \phi \geq 0$, and $\alpha + \phi \leq 1$

The dummy variable $D$ imposes a condition on the model that the news, regional events, economic turmoil have a strong and significant influence on the regional housing sector. The dummy value of $D$ is equal to one for $\varepsilon_{t-1} < 0$, when there is bad news that affect economy negatively and $D$ is zero otherwise. For instance, the regional business cycles tend to show a tranquil state when positive shocks influence the economies output and employment favorably. Using equation (2)-(4), we can define the residual vector $(\zeta)$ and the time varying conditional variance-covariance matrix $(\Sigma)$ as:

$$\zeta_t = \begin{bmatrix} \varepsilon_{i,t} \\ \varepsilon_{f,t} \end{bmatrix} = \begin{bmatrix} X_{i,t} - \beta_0^i - \beta_1^i\,X_{i,t-1} - \beta_2^i\,f_{t-1} \\ f_t - \beta_0^f - \beta_1^f\,X_{i,t-1} - \beta_2^f\,f_{t-1} \end{bmatrix} \tag{5}$$

$$\sum_t = \begin{bmatrix} \sigma_{i,t}^2 & \sigma_{if,t} \\ \sigma_{if,t} & \sigma_{f,t}^2 \end{bmatrix} = \begin{bmatrix} Var_t(X_t^i) & Cov_t(f_t X_t^i) \\ Cov_t(X_t^i f_t) & Var_t(f_t) \end{bmatrix} \qquad (6)$$

where $\sum_t$ is a symmetric and positive definite matrix, and $\sigma_{it,t} = \sigma_{fi,t}$; $\sigma_{it}^2 > 0$; $\sigma_{ft}^2 > 0$, and $\sigma_{it}^2 \sigma_{ft}^2 - \sigma_{if,t} \sigma_{fi,t} > 0$. For an estimated positive and statistically significant value of $\gamma$, the negative shocks have larger effects on $\sigma^2$ than positive shocks in equation (2)-(4), and their impact on the volatilities of housing sectors is also considered differently.

Estimation of the factor augmented bivariate T-GARCH model follows a two-step approach. In the first stage, the unobservable factors $f_t$ are estimated using the principal component analysis method, and in the second stage, the first of those factors $f_{1t}$ and the housing variable $X$ are jointly used for estimating T- GARCH model parameters through the following maximum likelihood function:

$$\log L(\theta) = -\frac{1}{2} \sum_{t=1}^n [\ln(2\pi) + \ln|\sum| + \zeta_t^T \sum_t^{-1} \zeta_t] \qquad (7)$$

$\theta$ denotes the vector of parameters in the GARCH models and $n$ is the monthly observation of housing variables and the macroeconomic factor.

The factor-GARCH specification originally developed by Engle (1987) and applied to the prediction of stock returns by Ng, Engle, and Rothschild (1992) as a multivariate application to volatility transmission by Lanne and Saikkonen (2007), and in bond returns by Engle et al. (1990). In recent REIT literature, Asteriou and Begiazi (2013) and Chang and Chen (2014) have utilized univariate GARCH models, Mondal (2014) examines volatility spillover using a bivariate GARCH model, and Begiazi, Asteriou, and Pilbeam, (2016) use constant mean M-GARCH to examine the volatility spillover of U.S. equity and mortgage REITs return. However, the model used in this study is a time-varying bivariate volatility model rather than the constant mean univariate-GARCH model. Moreover, this study's conditional volatility model considers the joint dynamics of macroeconomic factors and the housing variables. The choice of the GARCH model is motivated by the model's ability to capture unequal variances of the housing variables over time compared to any other alternative models, such as OLS,[6] auto-regressive integrated moving average (ARIMA), and auto-regressive conditionally heteroscedastic (ARCH) model. The standard GARCH model, as well as several parameterized extensions like exponential-GARCH (Nelson,1991; Nelson and Cao, 1992), GARCH-in-mean, threshold-GARCH (Glosten et al.,1993), and GARCH with exogenous factors (factor-GRACH), are among some of the most widely applied methods for modeling heteroscedastic variances and forecasting volatilities.

## 3.2 Housing risk factors and housing risk index

The individual housing risk factors are the conditional volatility of the ten housing variables derived from the factor augmented bivariate T-GARCH model above. Figure 3.1 shows the time series plots of the conditional volatility of the ten housing variables. A significant pattern in Figure 3.1 is that the individual housing volatility measures accurately match NBER's business cycle dates and, as expected, the volatility in the housing sector are higher during the 2007-8 financial crisis. The higher volatility in the measure of housing starts conditional volatility in four of the U.S. Census regions, and the mortgage contract interest rate is also consistent with the subprime mortgage crisis in 2008-09.

Similarly, volatility in real house prices, houses sold, and the ratio of houses for sales to houses sold is higher during 1980-83. The higher volatility in the early 1980s is related to the rise in the number of bank failures related to the savings and loan crises, which was continued steadily through 1982 because several regional and sectoral recessions hit financial and housing sectors. Since then, the volatility of all housing variables has declined significantly reached their historical lower values in early 2000.

Panel A of Table 3.1 presents the summary statistics of the ten housing variables for the sample period of 1972-2018. Panel B reports the correlation matrix for the estimated ten housing risk factors which shows estimated housing risk factors being positive and statistically significant, with maximum correlation of 0.92, and the minimum correlation of -0.03, and the average cross-correlations among housing risk factors are approximately 0.66. Panel C of Table 3.1 reports correlations between the standard macroeconomic and financial risk factors and housing risk factors. Most of the correlations are very low and statistically insignificant[7] with an average correlation of approximately 0.044, with a maximum of 0.46, and a minimum of -0.267. These lower and insignificant correlations suggest that the housing risk factors are independent from the conventional risk and uncertainty measures, and potentially capture a different dimension of the aggregate economy. Figure 3.2 shows the joint dynamics of the volatility of macroeconomic factor and the volatility of housing variables. There is a clustering of volatilities in housing series and the macroeconomic factor in all figures and the pattern shows that the clusters occur simultaneously. This pattern of volatility clustering rationalizes the choice of a bivariate GARCH model.

---

[6] The OLS is the natural choice to estimate how much a dependent variable change in response to a change in control variables. One of the OLS's main assumptions is homoskedasticity, which implies that the expected value of all error terms is constant over time. The homoscedasticity assumption regularly violates the OLS and requires considering alternative modeling in policy analysis.

Due to strong and persistent correlations among the conditional volatility of housing variables, common components estimated by principal component analysis can provide a realistic representation of the common comovement of all ten individual risk factors (Stock and Watson, 2016). In the current context, this correlation implies the possibility of constructing an alternative single measure of housing risk based on the existence of a common comovement across these ten measures and the assumption that this common factor is the primary source of volatilities in housing sector.

This new single volatility index or housing risk index is generated as linear combinations of PCA loadings[8] and the original ten volatility series. The first principal component in equation (8) explains a maximum of 78 percent of variation of these ten housing volatilities[9] and could be seen as a weighted sum of the volatilities of all ten housing variables.

$$
\begin{aligned}
House^{Risk\ Index} = \quad & 0.06\,\sigma_{HOUST} + 0.11\,\sigma_{PERMIT} + 0.01\,\sigma_{HPRICE} + 0.07\,\sigma_{HSTMW} \\
& + 0.03\,\sigma_{HSTNE} + 0.04\,\sigma_{HSTSOU} + 0.09\,\sigma_{HSTW} \\
& + 0.05\,\sigma_{HSOLD} + 0.98\,\sigma_{SARSOLD} + 0.04\,\sigma_{MORINT}
\end{aligned} \tag{8}
$$

---

[7] The level of significance is estimated using the corresponding p-values for all Pearson correlation coefficients. p < .001 '***', p < .01 '**', p < .05 '*'

[8] As Hastie, Tibshirani, and Friedman (2009) and Skinner, Holmes, and Smith (1986) show If $x_1, x_2, ..., x_n$ are the original series of $n$ number of correlated variables, then a weighted variable $y$ can be generated from a linear combination of these $n$ variables using $y = \lambda_1 x_1 + \lambda_2 x_2 + ... + \lambda_n x_n$ where the $\lambda_i$'s ($i = 1,2...n$) are the are the corresponding eigenvectors or principal component coefficients.

[9] PCA identifies how these 10 housing volatility series can be summarized in some fashion to construct a single index of meaningful aggregate volatility measures for further analysis and which captures maximum possible of the variation of these ten series.

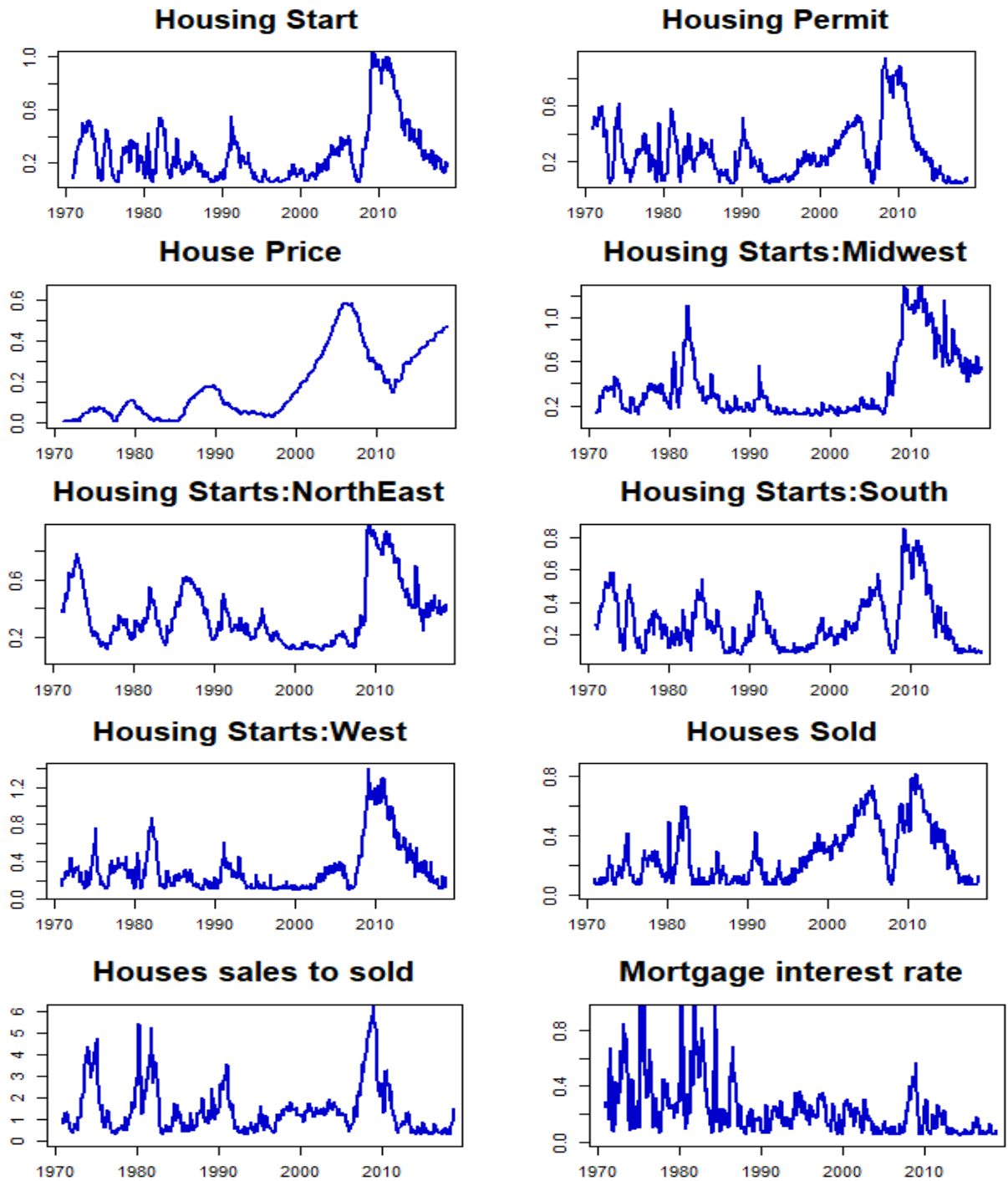

Figure 3.1. Conditional volatility of ten housing variables. For the sample period of January 1971- March 2018, conditional volatilities are estimated using the bivariate T-GARCH model with time varying parameter in equation (1) - (4).

Table 3.1: Summary statistics and correlation matrix

*Panel A: Housing risk factors*

| Statistic | N | Mean | St. Dev. | Min | Pctl(25) | Pctl(75) | Max |
|-----------|-----|-------|----------|-------|----------|----------|-------|
| HOUST | 576 | 0.275 | 0.224 | 0.052 | 0.117 | 0.351 | 1.097 |
| PERMIT | 576 | 0.268 | 0.198 | 0.046 | 0.110 | 0.351 | 0.941 |
| HPRICE | 576 | 0.177 | 0.165 | 0.003 | 0.045 | 0.294 | 0.586 |
| HSTMW | 576 | 0.380 | 0.297 | 0.111 | 0.164 | 0.522 | 1.362 |
| HSTNE | 576 | 0.358 | 0.211 | 0.111 | 0.199 | 0.462 | 0.995 |
| HSTSOU | 576 | 0.270 | 0.168 | 0.077 | 0.134 | 0.365 | 0.853 |
| HSTW | 576 | 0.326 | 0.263 | 0.097 | 0.141 | 0.376 | 1.409 |
| HSOLD | 576 | 0.266 | 0.189 | 0.065 | 0.108 | 0.375 | 0.814 |
| SATSOLD | 576 | 1.480 | 1.153 | 0.330 | 0.644 | 1.723 | 6.358 |
| MORINT | 576 | 0.319 | 0.237 | 0.020 | 0.122 | 0.529 | 0.905 |

.

*Panel B: Correlation matrix of the housing risk factors*

| | HOUST | PERMIT | HPRICE | HSTMW | HSTNE | HSTSOU | HSTW | HSOLD | SATSOLD | MORINT |
|--------|--------|--------|--------|--------|--------|--------|--------|--------|---------|--------|
| HOUST | 1*** | | | | | | | | | |
| PERMIT | 0.53*** | 1*** | | | | | | | | |
| HPRICE | 0.16** | 0.16** | 1*** | | | | | | | |
| HSTMW | 0.84*** | 0.42** | 0.22** | 1*** | | | | | | |
| HSTNE | 0.79*** | 0.33* | -0.03 | 0.74*** | 1*** | | | | | |
| HSTSOU | 0.81*** | 0.59*** | 0.11* | 0.52** | 0.55** | 1*** | | | | |
| HSTW | 0.92*** | 0.58*** | 0.14* | 0.85*** | 0.72*** | 0.7*** | 1*** | | | |
| HSOLD | 0.61*** | 0.47*** | 0.43*** | 0.47*** | 0.24** | 0.61** | 0.64*** | 1*** | | |
| SATSOLD | 0.23** | 0.5*** | 0.07* | 0.2** | 0.13* | 0.24** | 0.36** | 0.27** | 1*** | |
| MORINT | 0.33** | 0.12* | 0.15** | 0.61*** | 0.35** | 0.08* | 0.39** | 0.17* | 0.05 | 1*** |

*Panel C: Correlation matrix of the housing risk factors with standard macro-finance factors; $p < .001$ '***', $p < .01$ '**', $p < .05$ '*'*

| | HOUST | PERMIT | HPRICE | HSTMW | HSTNE | HSTSOU | HSTW | HSOLD | SATSOLD | MORINT |
|--------|---------|---------|----------|---------|---------|----------|---------|----------|----------|---------|
| Mkt-RF | 0.0545 | -0.0553 | -0.0240 | 0.0384 | 0.0468 | 0.0407 | 0.0493 | 0.0351 | -0.0823 | 0.0611 |
| SMB | 0.0199 | 0.0730 | -0.00113 | 0.0177 | -0.0435 | 0.0136 | 0.0498 | 0.0557 | 0.0314 | 0.0160 |
| HML | -0.0383 | -0.0387 | -0.0655 | -0.0564 | -0.0494 | -0.00743 | -0.0412 | -0.00934 | -0.0367 | -0.0185 |
| MOM | -0.0954 | -0.0657 | -0.0617 | -0.0857 | -0.0874 | -0.116 | -0.128 | -0.0839 | -0.0631 | -0.0640 |
| CrdSpr | 0.433 | 0.417 | 0.235 | 0.510 | 0.339 | 0.286 | 0.527 | 0.397 | 0.363 | 0.295 |
| D12 | 0.223 | -0.0438 | 0.448 | 0.464 | 0.190 | -0.0435 | 0.218 | 0.228 | -0.136 | 0.401 |
| E12 | 0.224 | -0.0738 | 0.439 | 0.429 | 0.154 | -0.00372 | 0.200 | 0.336 | -0.223 | 0.381 |
| b/m | 0.0138 | -0.0251 | -0.532 | 0.00937 | 0.0170 | 0.0397 | 0.0526 | -0.209 | 0.206 | 0.266 |
| lty | -0.336 | -0.154 | -0.666 | -0.342 | -0.255 | -0.173 | -0.267 | -0.357 | 0.159 | 0.0584 |
| ntis | -0.0755 | -0.125 | -0.517 | -0.308 | -0.171 | 0.0812 | -0.157 | -0.0587 | -0.243 | -0.425 |
| svar | 0.0749 | 0.209 | 0.0883 | 0.107 | 0.0695 | 0.0693 | 0.142 | 0.106 | 0.326 | -0.0337 |
| dfy | 0.242 | 0.274 | -0.121 | 0.323 | 0.188 | 0.215 | 0.361 | 0.132 | 0.507 | 0.415 |
| Skew | -0.165 | -0.0173 | -0.203 | -0.237 | -0.155 | -0.0351 | -0.171 | -0.0847 | 0.0301 | -0.178 |

Mkt-RF: excess return; SMB: Small Minus Big; HML: High Minus Low; MOM: Monthly Momentum Factor; CrdSpr: Credit spread; D12: 12-month moving sums of dividends paid on the S&P 500 index; E12: 12-month moving sums of earnings; b/m: Book-to-Market Ratio; lty: longterm government bond yield; ntis: Net Equity Expansion; svar: Stock Variance; dfy: Default Yield Spread; Skew: Skewness of stock returns.

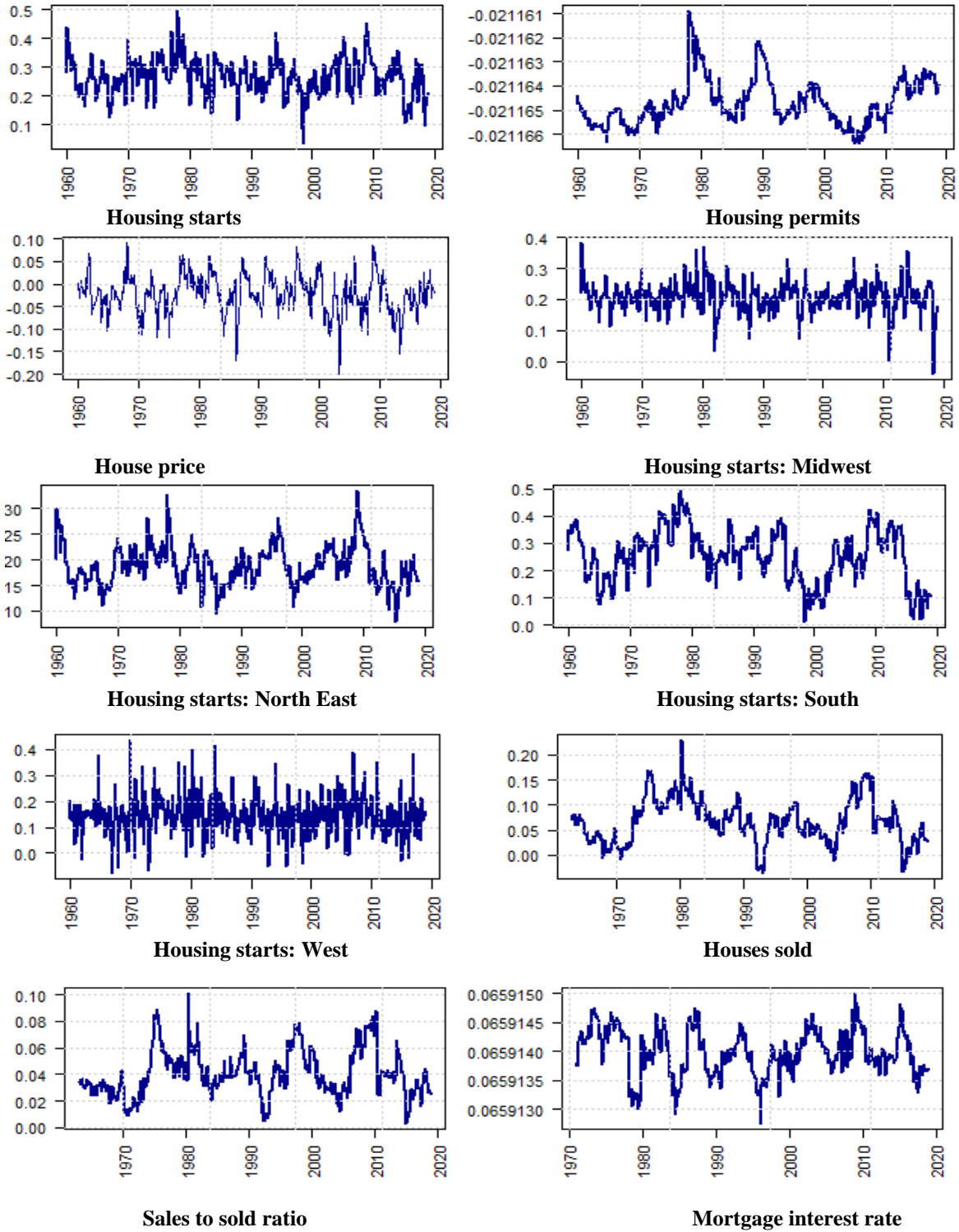

Figure 3.2: The joint dynamics of factor and the housing variables. The dynamics is measured using $\sigma_{x,f}$.

The construction of the proposed housing risk index in equation (8) is based on the factor loading matrix in Table 3.2. As shows in panel B of Table 3.1, volatility of all these ten housing variables in the correlation matrix are positively correlated with each other. Therefore, all the loadings in the first principal component are also positive in Table 3.2. The equation shows that the risk index has a higher weight (also refer as higher factor loadings) of 0.98 on the conditional volatility of the ratio of the house for sales to houses sold, whereas relatively smaller loadings of 0.01 on the conditional volatility of house price index. Housing permit has the second highest optimal weight, whereas conditional volatility of four housing starts variables have a sizeable role in constructing the housing risk index.

The proportion of variation explained by principal components in Table 3.2 shows that the first principal component explains the maximum 78 percent of overall variance and provides a good proxy for aggregate housing risk. The time series plot of the newly developed aggregate housing risk index in Figure 3.3 reflects similar patterns and fluctuations in individual housing risk factors in Figure 3.1. The proposed measure of aggregate housing risk is higher during 1982-85, 1990-92, and 2007-10 corresponding to the periods of the 1979 energy crisis, tight monetary policy, inflation-controlled recession, 1990 oil price shock, and the 2007-09 subprime mortgage crisis. The housing risk index also perfectly follows the NBER's business cycle fluctuations of peak and trough periods.

## Table 3.2: Summary of Principal Component Analysis

*Panel A: Importance of components*

| Statistic | PC1 | PC2 | PC3 | PC4 | PC5 | PC6 | PC7 | PC8 | PC9 | PC10 |
|---|---|---|---|---|---|---|---|---|---|---|
| Standard deviation | 1.17 | 0.48 | 0.23 | 0.18 | 0.15 | 0.13 | 0.11 | 0.084 | 0.075 | 0.044 |
| Proportion of Variance | 0.78 | 0.13 | 0.03 | 0.018 | 0.013 | 0.008 | 0.007 | 0.004 | 0.003 | 0.001 |
| Cumulative Proportion | 0.78 | 0.91 | 0.94 | 0.96 | 0.97 | 0.98 | 0.99 | 0.996 | 0.999 | 1.00 |

*Panel B: Factor loadings*

| Statistic | PC1 | PC2 | PC3 | PC4 | PC5 | PC6 | PC7 | PC8 | PC9 | PC10 |
|---|---|---|---|---|---|---|---|---|---|---|
| r1 | 0.055 | -0.429 | 0.057 | -0.136 | 0.011 | 0.184 | -0.093 | -0.250 | 0.083 | 0.823 |
| r2 | 0.091 | -0.177 | -0.208 | -0.410 | -0.497 | -0.652 | -0.144 | 0.226 | -0.022 | 0.057 |
| r3 | 0.011 | -0.069 | -0.544 | 0.251 | 0.310 | -0.061 | -0.668 | 0.015 | 0.292 | -0.048 |
| r4 | 0.065 | -0.546 | 0.105 | 0.491 | 0.189 | -0.427 | 0.120 | -0.210 | -0.386 | -0.132 |
| r5 | 0.031 | -0.347 | 0.300 | 0.201 | -0.369 | 0.411 | -0.384 | 0.503 | -0.149 | -0.135 |
| r6 | 0.041 | -0.247 | -0.051 | -0.461 | -0.094 | 0.264 | -0.232 | -0.593 | -0.177 | -0.451 |
| r7 | 0.094 | -0.472 | 0.037 | -0.077 | 0.067 | 0.037 | 0.361 | 0.116 | 0.731 | -0.274 |
| r8 | 0.049 | -0.223 | -0.394 | -0.357 | 0.450 | 0.194 | 0.234 | 0.450 | -0.406 | 0.003 |
| r9 | 0.984 | 0.153 | -0.003 | 0.063 | -0.007 | 0.053 | 0.009 | -0.023 | -0.02 | 0.009 |
| r10 | 0.042 | 0.045 | 0.629 | -0.341 | 0.517 | -0.268 | -0.347 | 0.133 | 0.061 | -0.030 |

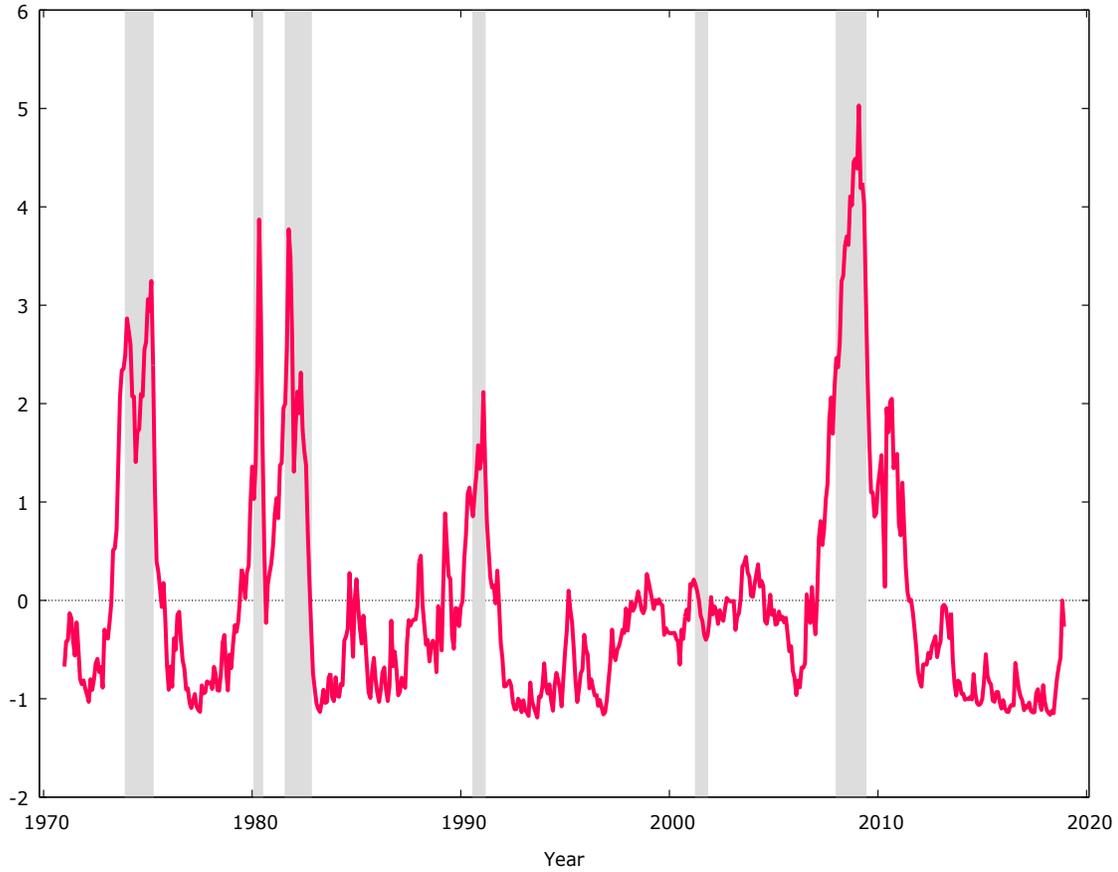

Figure 3.3: Aggregate Housing Risk Index. The index is defined as the first principal component of the volatilities of the ten housing variables explained in panel B of Table 3.1. The shaded area represents the NBER recession periods. The sample period is January 1971-December 2018.

## 4. Estimating the REIT returns across various asset pricing theories

### 4.1 The Arbitrage Pricing Model

Cox et al. (1985) and Ross (1976) propose a macro-factor based arbitrage Pricing Theory (APT) model that incorporates a set of macroeconomic variables, with the assumption that returns $r$ on $N - assets$ are related with a number of $k$ factors as in the following model:

$$r_{Nx1} = \mu_{Nx1} + \beta_{Nxk}\ f_{kx1} + \varepsilon_{Nx1} \tag{9}$$

where, $\mu_{Nx1}$ is vector of expected returns, $f_t \sim N(0,1)$ is a factor matrix, $\beta_{Nxk}$ is a factor loading matrix, and $\varepsilon_{Nx1}$ is a vector of residuals. If the assumption of exact pricing condition holds and if there is a risk-free asset and its return is $r_f$, then the expected returns are given by

$$\mu_{Nx1} = I\, r_f + \beta_{1,Nxk}\,[.]_{kx1} \tag{10}$$

where, $I$ is a vector of ones, and $[.]$ is vector of constant prices of risk. Using equation (9) and (10)

$$r_{Nx1} = I\, r_f + \beta_{1,Nxk}\,[.]_{kx1} + \beta_{Nxk}\,f_{kx1} + \varepsilon_{Nx1}$$
$$r_{Nx1} - I\, r_f = \beta_{1,Nxk}\,[.]_{kx1} + \beta_{Nxk}\,f_{kx1} + \varepsilon_{Nx1}$$
$$ER_{Nx1} = \beta_{1,Nxk}\,[.]_{kx1} + \beta_{Nxk}\,f_{kx1} + \varepsilon_{Nx1} \tag{11}$$

where $ER_{Nx1}$ is a vector of excess return. Equation (11) suggest that the excess return of $N-assets$ are generated by the constant risk premium and the covariance of returns with the random factors. Intuitively, the APT model shows that the variation of excess return can be decomposed into the contribution related to constant risk premium of the each factors in $[.]$ and the covariance of return to the set of factors $f_{kx1}$. Risk associated with the expected asset returns in the APT model are derived from two sources of risk factors. One is macroeconomic and financial risk factors that are non-diversified and the second source is asset specific risk that can be diversified.

The factors $f_{kx1}$ in equation (11) includes macroeconomic factors and are not directly estimated by the APT. Given that the APT offers limited information to investors when they attempt to identify macroeconomic risk factors and their expected risk premia, previous studies have focused on a set of observed macroeconomic and financial variables and their volatility as candidates for risk factors. For example, Cox et al. (1985) estimate factors using an intertemporal asset pricing model where factors are "multivariate" proxy of the arbitrary set of observable and unobservable macroeconomic variables. For U.S. stock returns, Chen (1983) and Chen, Roll, and Ross (1986) find the volatility of real industrial production, yield curve, inflation, and other sector-specific macroeconomic variables are statistically significant observed risk factors. Following Chen et al. (1986), a plethora of empirical tests of the APT (Ahamed, 2021; Antoniou, Garrett, & Priestley, 1998; Burmeister and McElroy, 1988; Connor and Korajczyk, 1988; French, 2017; Priestley, 1996; Shukla and Trzcinka, 1990) also propose a range of macroeconomic variables as common risk factors, with the confirmation that expected asset returns are significantly related to observed macroeconomic and financial factors.

In the spirit of the APT model and following Chen (1983) and Chen et al. (1986), this study estimates a single risk factor from a set of housing variables and argues that the proposed housing risk factor is a proxy for a sector-specific macroeconomic risk factor in equation (11). This factor can explain the time-series behavior of expected REIT returns because it captures the dynamics and the pervasive information of the volatility of the U.S. housing sector. According to the APT model, regressing expected REIT equity returns on the housing risk factor follows the specification of the linear factor model, and the estimated coefficients should be positive.[10] This positive relationship provides the theoretical support to incorporate the housing risk factor to evaluate the empirical performance of the APT model.

## 4.2 The Intertemporal Capital Asset Pricing Model (ICAPM)

The APT model suggests a positive relationship between all risk variables and the expected REIT equity returns. However, macroeconomic factors in the APT model are non-diversifiable. Also, the factor-based model's statistical nature does not provide any economic intuition of why factors deserve compensation. A further investigation of the equilibrium risk-return relationship in Merton's (1973) ICAPM validates the role of aggregate housing risk index in predicting REIT equity return. The ICAPM shows the following relationship:

$$E(R_{m,t}|I_{t-1}) = \alpha \, \text{var}(R_{m,t}|I_{t-1}) + \beta \, \text{cov}(R_{m,t}, X_t|I_{t-1}) \tag{12}$$

where, $R_{m,t}$ is the conditional excess return on the market portfolio $m$ at the time $t$, $I_{t-1}$ denotes the available information set up to the time $t-1$ that investors use to make a prediction about the expected excess returns $E(R_{m,t}|I_{t-1})$ at time $t$, $\text{var}(R_{m,t}|I_{t-1})$ is the conditional variance of the excess returns of market portfolio., and $\text{cov}(R_{m,t}, X_t|I_{t-1})$ measures the conditional covariance between the excess returns on market portfolio and the state variables $X$ at time $t$. The innovations

---

in the set of state variables affect the investment opportunities at time $t$ conditioned on the available information set at time $t-1$. Thus, the equation states that the expected returns on an asset can be divided into compensation for bearing the risk related to the market and the risk generated from the set of the state variable $X$. So,

$$E[\beta_{mx,t} \mid I_{t-1}] = \frac{\text{cov}[R_{m,t} X_t \mid I_{t-1}]}{\text{var}[X_t \mid I_{t-1}]} \tag{13}$$

The proposed housing risk index developed in section (3) can be considered a proxy measure of a factor in the set of state variables $X$ in equation (13) and the estimated beta can define as the housing risk beta. In the ICAPM specification, volatility in the housing sector is then a source of non-diversifiable systematic risk. Therefore, the model predicts that the investors holding REIT portfolios must be compensated by a risk premium proportional to the degree of housing risk exposure.

Campbell (1996) also provides a multifactor intertemporal asset pricing model where asset returns are related to the covariance of the market returns and the innovations in the future investment opportunity set. In the context of asset risk premia, a risk-averse investor must compensate by the risk premium, which leads to a positive correlation of asset returns with the innovations in the future investment opportunity set. Recent theoretical (Anderson, Ghysels, & Juergens, 2009; Bekaert, Engstrom and Xing, 2009; Bloom, Bond, and Reenen, 2007; Pramanik, 2020; Polansky and Pramanik, 2021) and empirical studies (Bali et al. 2014; Caldara, Fuentes-Albero, Gilchrist, & Zakrajšek, 2016) also provide evidence that the volatility in macroeconomic variables can be a relevant proxy measurement of risk in investor's investment opportunity set. Therefore, an increase in macroeconomic volatility implies a reduction of asset return through an unfavorable shift in the investment set, and investors require a higher expected return for bearing higher risk.

Overall, the ICAPM model provides a theoretical foundation that the proposed housing risk factor is a relevant risk factor in REIT investor's investment set, with the assumption that the estimated beta should positively predict the future REIT equity returns. So, equation (13) suggests that the housing risk index is a proxy measure of broader macroeconomic factor in state variable set $X$, and the expected market risk premium $E[\beta_{mx,t} \mid I_{t-1}]$ at time $t$ contains compensation for the risk in the U.S. housing sector.

# 5. Analysis of results

## 5.1 Data

The data set consists of monthly return on three REIT equity portfolios spanning from January 1971 to December 2018. There are 576 monthly observations for the two REIT equity portfolios and 228 observations for the real estate equity return. The REIT equity portfolio data is from the U.S. national association of real estate investment trusts (NAREIT) website. The site continuously tracks the information on the performance of the U.S. REIT trading and quotations of securities. The Fama-French-Carhart four factors are from the Fama-French research website. The risk-free rate is the three-month treasury bill rate taken from the St. Louis Federal Reserve Bank website, the remaining U.S. economic and financial predictors are from Goyal's website, and finally, the average of monthly skewness of U.S. firm-level return is from Jondeau, Zhang, and Zhu (2019). The construction of economic and financial data, a complete list of macroeconomic variables and their sources are in Appendix I and II.

Panel A of Table 3.3 provides summary information about REIT equity monthly returns and related statistics. The summary statistics of standard macro-finance predictive variables are in panel B. All three REIT indexes show a positive mean return value over the entire sample period. The REIT equity return index has the highest average return of 1.033, and the higher return accompanied by the lowest standard deviation of 4.86. The distribution of the monthly REIT returns shows fat tails and nonzero skewness coefficients. The Jarque-Bera (JB) test suggests that the normality assumption of return distribution holds in all REIT returns. Finally, a portmanteau Q-test of order one rejects the presence of autocorrelation.

Table 3.3: Summary statistics

*Panel A: Descriptive statistics of REIT equity index monthly returns*

| REIT Index | N | Mean | Std. Dev. | Min | Max | Skew | Kurt | JB | Q (1) |
|---|---|---|---|---|---|---|---|---|---|
| Total return | 576 | 0.878 | 5.01 | -30.23 | 30.81 | -0.387 | 10.38 | 1293.5*** | 1.88 |
| Real estate50 | 228 | 1.025 | 5.80 | -30.95 | 28.05 | -0.92 | 9.83 | 476.91*** | 0.21 |
| Equity return | 576 | 1.033 | 4.86 | -31.67 | 31.02 | -0.67 | 10.86 | 1494.1*** | 2.71 |

*Panel B: Summary statistics of predictive variables*

| Variables | N | Mean | Std. Dev. | Min | Max | Skew | Kurt |
|---|---|---|---|---|---|---|---|
| Housing Risk Index | 576 | 0.000 | 2.25 | -2.62 | 8.34 | 1.8 | 3.05 |
| Market excess return | 576 | 0.54 | 4.47 | -23.24 | 16.1 | -0.55 | 2.08 |
| SMB | 576 | 0.15 | 3.06 | -16.86 | 21.7 | 0.56 | 6.23 |
| HML | 576 | 0.31 | 2.89 | -11.18 | 12.87 | 0.09 | 2.01 |
| Risk free rate | 576 | 0.38 | 0.28 | 0.000 | 1.35 | 0.61 | 0.34 |
| MOM | 576 | 0.65 | 4.31 | -34.33 | 18.36 | -1.37 | 10.69 |
| Credit spread | 576 | 2.22 | 0.71 | 0.93 | 6.01 | 1.39 | 4.31 |
| D12 | 576 | 16.45 | 12.37 | 3.07 | 53.74 | 1.19 | 0.74 |
| E12 | 576 | 39.02 | 32.28 | 5.16 | 132.39 | 0.98 | -0.22 |
| b/m | 576 | 0.48 | 0.28 | 0.12 | 1.20 | 0.87 | -0.47 |
| lty | 576 | 0.067 | 0.028 | 0.017 | 0.14 | 0.46 | -0.21 |
| ntis | 576 | 0.007 | 0.02 | -0.05 | 0.045 | -0.54 | -0.11 |
| svar | 576 | 0.002 | 0.004 | 0.0001 | 0.07 | 9.7 | 120.0 |
| dfy | 576 | 0.01 | 0.004 | 0.005 | 0.033 | 1.79 | 4.19 |
| Skew (U.S. stock return) | 576 | 0.051 | 0.026 | -0.085 | 0.131 | -0.62 | 2.49 |

SMB: Small Minus Big; HML: High Minus Low; MOM: Monthly Momentum Factor; CrdSpr: Credit spread; D12: 12-month moving sums of dividends paid on the S&P 500 index; E12: 12-month moving sums of earnings; b/m: Book-to-Market Ratio; lty: long-term government bond yield; ntis: Net Equity Expansion; svar: Stock Variance; dfy: Default Yield Spread; Skew: Average skewness of U.S. firm level stock returns
*** indicates significance at the 1% level.

## 5.2 Baseline Univariate regression with housing risk index

The baseline univariate regression model is

$$R_t^i = \alpha^i + \beta^i H_{t-1} + \varepsilon_t^i \tag{14}$$

where $R_t^i$ is the monthly excess REIT returns at time $t$, in which $i$ represents one of the three REIT equity returns, $H_t$ is the housing risk index in month $t-1$, and $\beta^i$ is the housing risk beta. The Newey and West (1987) t-statistic is used to test the statistical significance of estimated coefficients.

The OLS and the quantile regression model test whether estimated coefficients of housing risk index satisfy standards risk-return relationship for forecasting the REIT equity return. Table 3.4 presents the OLS and quantile regression results. The first row shows that the OLS estimates of $\beta^i$ for all REIT returns are positive and statistically significant, suggesting that an increase in risk in housing sectors predicts risk premium for the REIT equity market.

Since the housing risk index is standardized, the estimated $\beta$ implies that a one standard deviation increase in housing risk factor predicts a 7.61 percent increase in all REIT excess return and a 6.01 percent for REIT equity excess return. Looking at historical REIT return data, the average annual returns of REIT lies between 7 percent and 8.6 percent over the 1977-2012 periods. For the period of 2012-17, the average annual all REIT equity total return is an average of 11.64 percent (REIT, 2018).

The adjusted $R^2$ values of the OLS regression model lies in between 4.02 percent and 5.6 percent. While the relatively smaller values of $R^2$ is mainly due to the absence of a large set of explanatory variables, the comparison of model performance in section 5.6 provides superior out-of-sample forecasting performance of univariate and multivariate models. Next, following Huang, Jiang, Tu, and Zhou (2015), I further investigate the predictive power of housing risk beta by adding NBER recession dummy and the Chicago Board of Options Exchange's (CBOE) implied volatility index (VIX) to the equation (14):

$$R_t^i = \alpha^i + \beta^i H_{t-1} + NBER_t + \varepsilon_t^i$$
$$R_t^i = \alpha^i + \beta^i H_{t-1} + VIX_t + \vartheta_t^i$$

(14.1)

where NBER is a dummy variable that takes a value one for NBER recession and zeroes for expansion. The VIX index is a popular measure of aggregate market volatility that commonly referred as the forward looking barometer of market's risk. Panel A of Table 3.4 reports the

Table 3.4: Baseline regression with housing risk index

*Panel A: OLS regression*

================================================================================

| | | Dependent variable | |
| | | --- | |
| Housing beta | All REIT | Real Estate50 Equity | All REIT Equity |

--------------------------------------------------------------------------------

| | | | |
|---|---|---|---|
| Univariate OLS | 0.076* | 0.071* | 0.061* |
| | (0.04) | (0.06) | (0.04) |
| Adj R-Sq | 5.6 | 4.3 | 4.1 |

--------------------------------------------------------------------------------

| OLS (with NBER dummy) | 0.096** | 0.086** | 0.063* |
| Adj R-Sq | 1.5 | 1.5 | 0.4 |
| N | 576 | 228 | 576 |

--------------------------------------------------------------------------------

| OLS (with VIX) | 0.155** | 0.152* | 0.153** |
| Adj R-Sq | 7.8 | 8.1 | 7.4 |
| N | 332 | 228 | 332 |

================================================================================

*Panel B: Univariate quantile regression*

================================================================================

| | | Dependent variable | |
| | | --- | |
| Housing beta | All REIT | Real Estate50 Equity | All REIT Equity |

--------------------------------------------------------------------------------

| | | | |
|---|---|---|---|
| Q1 (0.25) | -0.03 | -0.05 | -0.07 |
| | (0.04) | (0.12) | (0.05) |
| Q2 (0.50) | 0.15*** | 0.16*** | 0.13*** |
| | (0.04) | (0.06) | (0.04) |
| Q3 (0.75) | 0.19*** | 0.18*** | 0.17*** |
| | (0.04) | (0.06) | (0.04) |
| Q4 (0.95) | 0.30*** | 0.38*** | 0.39*** |
| | (0.07) | (0.07) | (0.07) |
| N | 576 | 228 | 576 |

================================================================================

| Note: | *p<0.1; **p<0.05; ***p<0.01 |

regression results. As shown in column (1)-(3), $\beta^i$'s are positive and economically meaningful for all REIT portfolios. Like the baseline univariate regression, a one standard deviation increase in housing risk predicts a maximum of 9.6 percent excess return, implying that the housing risk factor is a state variable in investors' investment opportunity sets. The estimated housing risk $\beta^i$'s are also positive and statistically significant when models include NBER dummy and CBOE's volatility index.

In addition to the conditional mean forecast, the empirical examination also employs the quantile regression approach (Koenker and Bassett, 1978) that focuses on the REIT return predictability at different quantiles of the return distribution. The approach examines the relationship between REIT excess returns and the housing risk for the full range of distribution of returns. Compared to the least-squares regression model, the quantile estimation removes the impact of the left- and right-tailed error distribution and can provide a robust estimation of housing risk betas (Chiang and Li, 2012; Hua, Polansky, & Pramanik, 2019). While both the univariate and multivariate regression model considers a wide range of quantiles, for the brevity of space, Table 3.4 only reports 25%, 50%, 75%, and 99% points of the return distribution.

Consistent with results in Panel A, the estimated betas are positive and significant for the quantiles above 10 %, providing evidence that the housing risk beta significantly predicts risk premiums over the various distributions points. Among the four quantile estimates of $\beta^i$, 50%, 75%, and 99% quantile beta are highly significant, where the $\beta^i$'s are much higher in higher quantiles. This finding is also consistent with Chiang and Li (2012), who also finds higher positive betas above the median quantile of the distribution of returns.

## 5.3 Housing risk beta with control variables

The baseline OLS and quantile regression findings show a strong positive risk-return relation and the economically meaningful predictive ability of the housing risk index. This section investigates whether the risk-return relationship of housing beta remains unchanged when regression models include standard control variables widely used in asset price literature.

In financial economics, controlling Fama-French three factors and Carhart's momentum factor (MOM) are the standards for predicting conditional mean returns (Fama and French, 1989; Ferson and Harvey, 1991). Following Welch and Goyal (2008), recent empirical studies (Ahamed, 2021; Alam, Khondker, & Molla, 2013; Bollerslev, Marrone, Xu, and Zhou, 2014; Goh, Jiang, Tu, and Wang, 2013; Hossain and Ahamed, 2015; Pramanik and Polansky, 2021; Rapach, Strauss, and Zhou, 2013) also use a set of financial and macroeconomic variables including the book-to-market ratio, earnings-price ratio, the dividend yield default premia, yield spread between BAA and AAA bonds, short and long interest rates, stock variance, net equity expansion, consumer price index, and inflation to investigate the risk-return hypothesis. Also, Conrad, Dittmar, and Ghysels (2013), Harvey and Siddique (2000), and Jondeau et al. (2019) highlight the predictive power of conditional skewness of U.S. stock returns for predicting equity risk premium.

The multivariate regression specification adds twelve control variables to the baseline regression model: the Fama-French-Carhart four factors, seven U.S. economic and financial predictors from Goyal's website, and the average of monthly skewness of Jondeau et al. (2019). However, many of the standard predictor variables found to be statistically insignificant (as also document by Cenesizoglu and Timmermann (2007) and Welch and Goyal (2008)) for predicting the left and right tail of the return distribution. While adding more predictors provide better goodness of fit of the data to the model, too many predictors also penalize models by overfitting the data. Also, Campbell (1987) and Ghysels, Santa-Clara, and Valkanov (2005) suggest that the non-robust estimation of a positive risk-return relationship in many empirical studies is partly due to lack of guidance of the selection of predictive variables in the multivariate asset price models.

A Bayesian variable selection approach is applied to overcome this limitation, which compares model performance using log posterior odds and selects the model's covariates accordingly. In general, the approach selects the best set of predictors and suggests the best model with the smallest log posterior odds. There are twenty competing models for each predictor, where all models are ordered according to the log of posterior probabilities. The selection of predictors starts with the model that considers the full set of twelve predictors. Next, the approach uses the log posterior odds values to compare alternative models with selected predictors.

Figures 3.4-3.6 shows the competing models, along with their log-odds and selected predictors. In each figure, twelve predictors and the intercept are on the $y$-axis, while the values

of log posterior odd corresponding to each model are on the *x*-axis. The black blocks represent the predictors that are not included in a model. The color of red, dark-red, and purple of each column corresponds to the model's log of the posterior probability. Models are clustered together with the same colors and have the same posterior probabilities. If we view models by rows, various colors show whether a variable is included in a specific model. For example, Mkt-RF, SMB, and HML appear in all twenty models to predict all REIT total returns in figure 3.4. The figure also shows that the Bayesian approach ranks all competing models and selects Mkt-RF, SMB, HML, MOM, E12, ntis, and dfy predictors[11] for the best model that results from the lowest log posterior odds of 0.045.

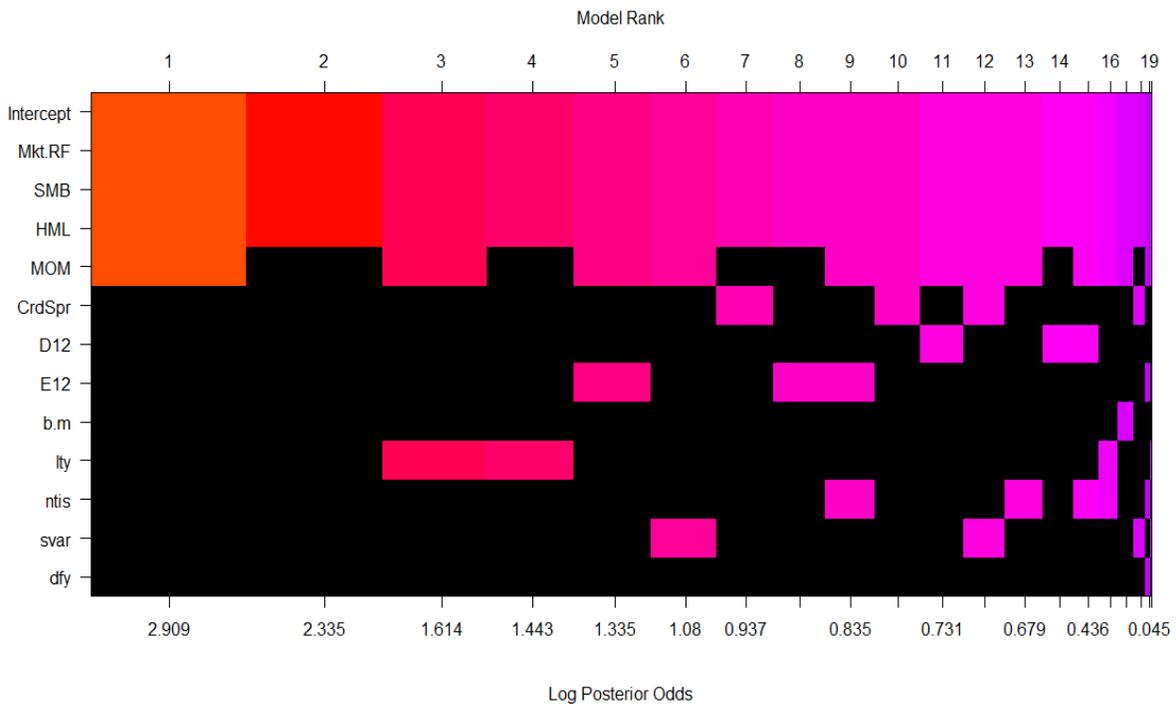

Figure 3.4: Selection of covariates: All REIT total return

---

[11] SMB: Small Minus Big; HML: High Minus Low; MOM: Monthly Momentum Factor; CrdSpr: Credit spread; D12: 12-month moving sums of dividends paid on the S&P 500 index; E12: 12-month moving sums of earnings; b/m: Book-to-Market Ratio; lty: long-term government bond yield; ntis: Net Equity Expansion; svar: Stock Variance; dfy: Default Yield Spread; Skew: Average skewness of U.S. firm-level stock returns

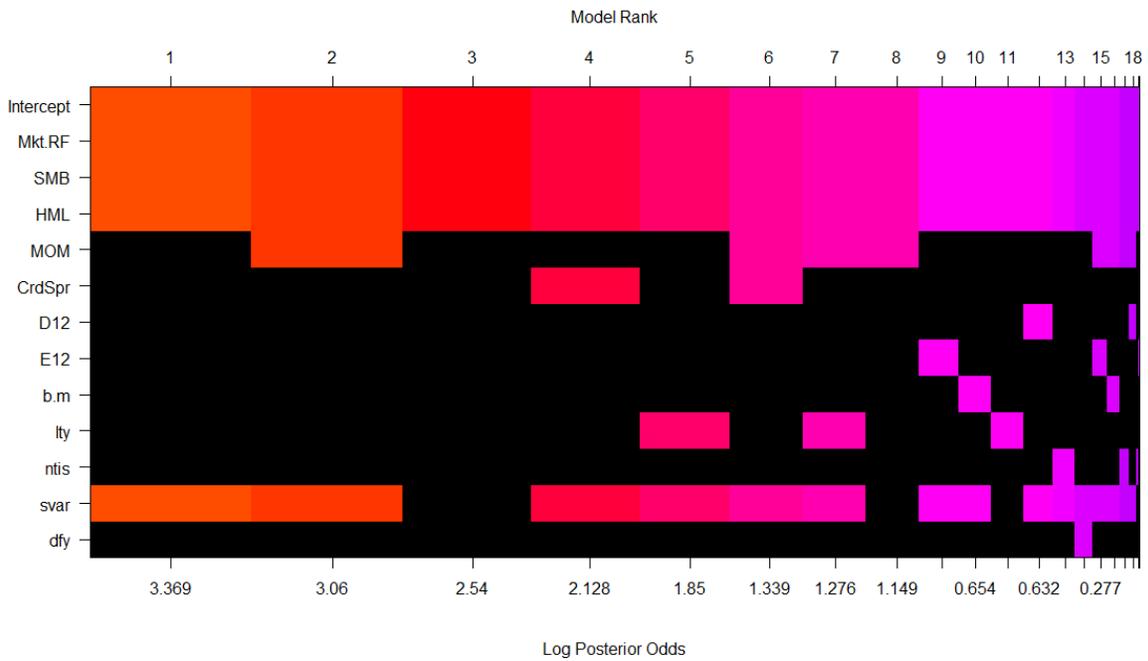

Figure 3.5: Selection of covariates: Real Estate50 equity return

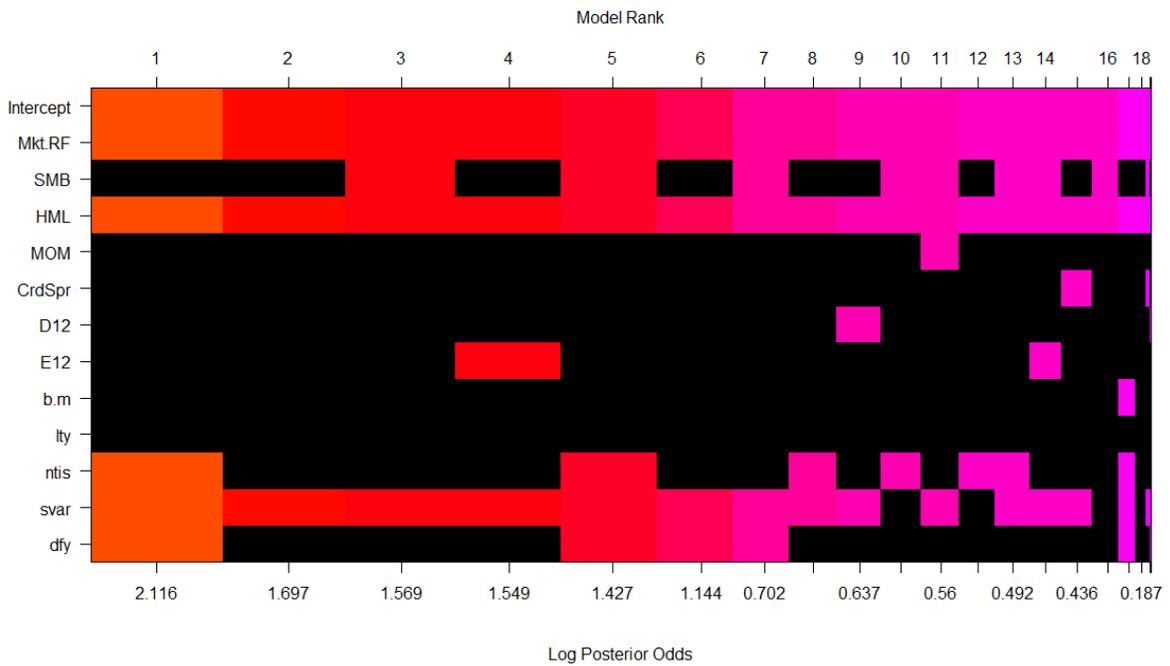

Figure 3.6: Selection of covariates: All REIT equity return

Panel A of Table 3.5 reports the multivariate OLS regression results, where the Bayesian log posterior odds select a relevant set of model covariates. After controlling for a selected predictor, the housing risk beta remains positive and statistically significant for the three REIT portfolios. While the estimated housing beta for all REIT total return and equity returns are smaller than the univariate estimates, statistically significant beta at 5 percent level of significance suggests the robust predictive ability of the housing risk index. The $R^2$-values of all three models are higher than the univariate models, ranging from 48 percent to 55 percent, suggesting the validity of inclusion of housing risk index with the standard predictors in the models.

Panel B of Table 3.5 provides evidence for a positive and significant housing risk beta for the 50%, 75%, and 99% quantile of the return distribution. After controlling for the same set of predictors, the housing risk index appears to be highly significant for the quantiles above 50 % of the return distribution. For the real estate50 equity return, housing risk beta remains insignificant for most of the quantiles. Consistent with the univariate regression, the quantile coefficient of housing risk betas are also relatively higher for the higher quantile of the return distribution.

## 5.4 Sub-sample analysis

The robustness analysis of the results focuses on the housing risk index's predictive ability obtained from the OLS and quantile estimation of the Fama-French factor model. This section re-estimates the Fama-French three-factor model using two sub-samples to examine if the estimated housing risk beta is sensitive to the sample periods. The estimation assumes multiple unknown structural breakpoints over the sample periods and applies the endogenous identification approach proposed by Bai and Perron (1998). The approach detects four corresponding breakpoints in figure 3.7: 1984:M3, 1993:M7, 2002:M4, and 2009:M4. The estimation of the model using structural breaks allows the study to examine the cases if the risk-return relationship changes due to an expected policy change.

Table 3.5: Housing risk beta with standard control variables

*Panel A: OLS regression with the standard control variables.*

==================================================================================

| | Dependent variable | | |
| Predictors | All REIT | Real Estate50 Equity | All REIT Equity |
|---|---|---|---|
| Housing risk index | 0.028** | 0.019* | 0.026** |
| | (0.01) | (0.017) | (0.01) |
| Mkt-RF | 0.63*** | 0.54*** | 0.57*** |
| | (0.03) | (0.05) | (0.03) |
| SMB | 0.24*** | | 0.21*** |
| | (0.02) | | (0.03) |
| HML | 0.35*** | | 0.32*** |
| | (0.03) | | (0.03) |
| MOM | −0.08** | | -0.06** |
| | (0.03) | | (0.03) |
| dfy | -0.0001 | | |
| | (0.03) | | |
| ntis | 0.05* | | |
| | (0.02) | | |
| E12 | 0.12** | | |
| | (0.03) | | |
| svar | | -0.16*** | -0.11*** |
| | | (0.05) | (0.03) |
| Adj R-Sq | 0.55 | 0.48 | 0.51 |
| N | 576 | 228 | 576 |

==================================================================================

*Panel B: Quantile regression with the standard control variables*

==================================================================================

| | Dependent variable | | |
| Housing beta | All REIT | Real Estate50 Equity | All REIT Equity |
|---|---|---|---|
| Q1 (0.25) | 0.0008 | 0.007 | -0.006 |
| | (0.014) | (0.023) | (0.013) |
| Q2 (0.50) | 0.043*** | 0.028 | 0.033*** |
| | (0.013) | (0.017) | (0.011)) |
| Q3 (0.75) | 0.05*** | 0.014 | 0.056*** |
| | (0.014) | (0.013) | (0.014) |
| Q4 (0.95) | 0.027*** | 0.003 | 0.067*** |
| | (0.016) | (0.022) | (0.026) |
| N | 576 | 228 | 576 |

==================================================================================

Note: *p<0.1; **p<0.05; ***p<0.01

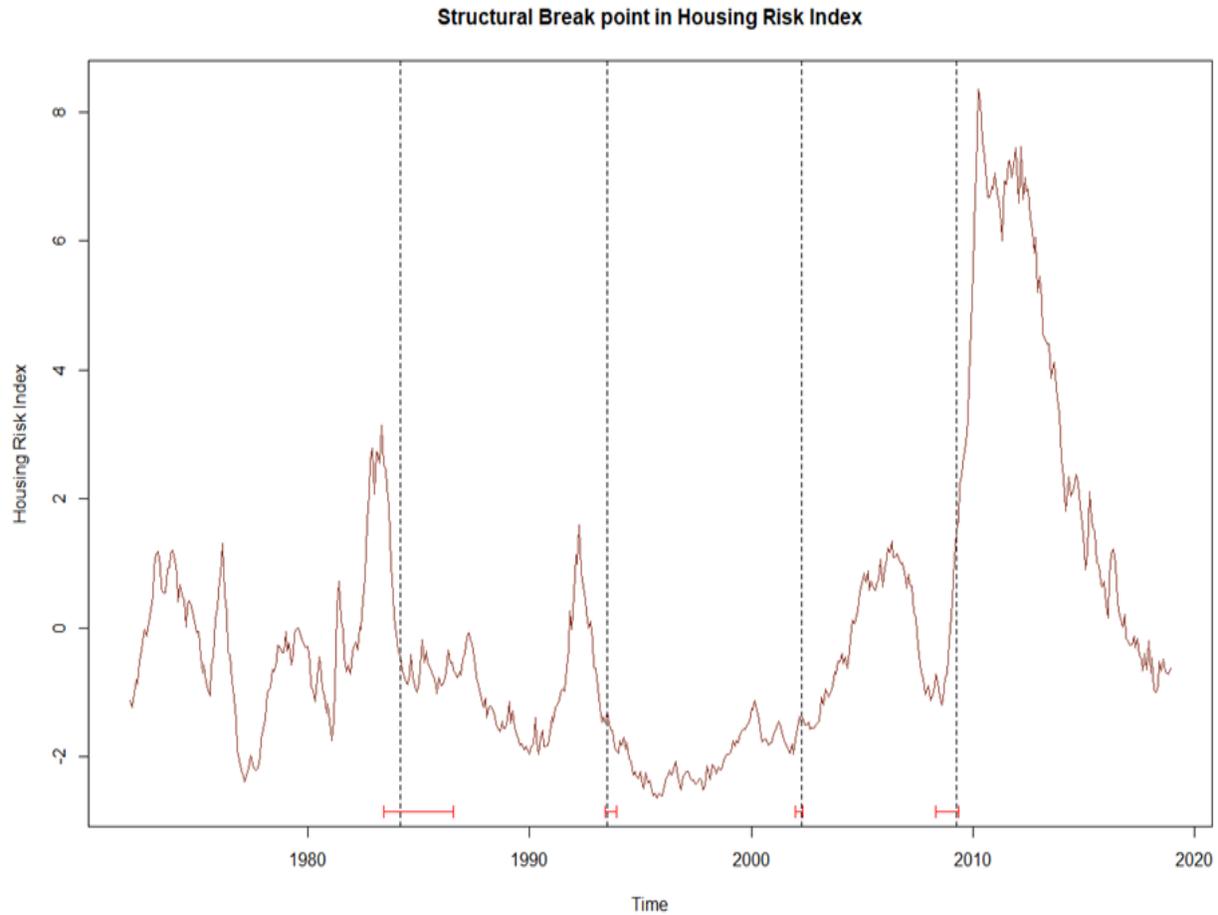

Figure 3.7: Endogenous identification of structural breakpoints. Red color horizontal bands indicate the breakpoints

Table 3.6 reports the estimated housing risk beta from the OLS and quantile estimation of the Fama-French model. The results are qualitatively similar for all breakpoints when the sample is divides into four sub-group based on the four endogenous breakpoints. Table 3.6 reports one subsample results using the 2009:M4 structural breakpoint.

On the prediction of the positive relationship between the housing risk index and the REIT returns, the results are similar to those of full sample analysis reported in Table 3.6. However, the OLS model results show that the negative shock in the 2007-09 financial crisis tends to require higher REIT returns as a risk premium for the post-crisis periods. The estimated housing risk beta is 10.8 for the pre-financial crisis periods, while the average beta value is 17.2 for the 2009-18

## Table 3.6: Housing risk beta in sub-samples

| | | Dependent variable | | |
|---|---|---|---|---|
| Sub-samples | Model | All REIT total | Real estate50 | All REIT equity |
| 1971M1-2009M4 | OLS | 0.108** | 0.12 | 0.102** |
| | | (0.047) | (0.09) | (0.04) |
| | Adj R-sq | 0.22 | 0.23 | 0.34 |
| | N | 447 | 112 | 447 |
| 1971M1-2009M4 | Q1 | 0.11** | -0.34** | 0.067* |
| | | (0.04) | (0.14) | (0.035) |
| | Q2 | 0.063 | 0.14*** | 0.085** |
| | | (0.04) | (0.05) | (0.041) |
| | Q3 | 0.067* | 0.49*** | 0.084** |
| | | (0.03) | (0.06) | (0.03) |
| | Q4 | 0.09 | 1.33*** | 0.23*** |
| | | (0.06) | (0.14) | (0.06) |
| 2009M5-2018M12 | OLS | 0.172* | 0.174* | 0.171* |
| | | (0.092) | (0.09) | (0.09) |
| | Adj R-sq | 0.14 | 0.13 | 0.14 |
| | N | 116 | 116 | 116 |
| 2009M5-2018M12 | Q1 | 0.004 | -0.0034 | -0.013 |
| | | (0.16) | (0.15) | (0.18) |
| | Q2 | 0.29** | 0.33*** | 0.24** |
| | | (0.09) | (0.10) | (0.11) |
| | Q3 | 0.35** | 0.26*** | 0.28*** |
| | | (0.10) | (0.05) | (0.06) |
| | Q4 | 0.38* | 0.403*** | 0.37** |
| | | (0.17) | (0.11) | (0.16) |

Note:  *: $p<0.1$; **: $p<0.05$; ***: $p<0.01$

periods, reflecting the effect of higher volatiles of U.S. housing sectors on the expected REIT returns. As the findings in previous sections, housing betas are positive and significant for the 50%, 75%, and 99% quantile of the return distribution. Therefore, the positive relationship between the housing risk index and REIT returns is not affected by choice of two sample periods, which validates the robustness of the findings of this study.

## 5.5 Out-of-sample forecast analysis

The out-of-sample forecasting analysis splits the total sample into a training set (in-sample) and a validation set (out-of-sample). Following Welch and Goyal (2008), the ratio of the training and validation sample is 80:20 of the total sample. The directional correlation accuracy, Mean Square of Forecast Error (MSFE), and the residual standard error evaluate the forecasting accuracy. The directional correlation accuracy is the percentage measure of the correlation between the actual values of REIT excess returns and the predicted values. A higher percentage of correlation suggests a more accurate out-of-sample predictive direction to the actual values' direction. The MSFE measures the quality of predictors by taking expected squared distance between model's predicted values and what the actual value of these predictors. The correlation accuracy is often considered as the most popular metric to the REIT investors because the method can detect the future directional movement of housing variables.

The analysis compares univariate, multivariate, and the Fama-French three-factor model to test the regression models' best forecasting ability. For one period ahead, forecasting of REIT excess returns, panel A of Table 3.7 compares the in-sample and out-of-sample regression results along with the values of forecasting evaluation matrices. Considering the directional accuracy, panel A reveals that the highest in-sample accuracy achieved with the Fama-French three-factor model followed by the multivariate model with Welch and Goyal (2008) risk factors. Additionally, the residual standard error is 0.66 for the Fama-French model while that of the multivariate model with standard risk factors is 1.66. Therefore, if the correlation accuracy and the residual standard

Table 3.7: In-sample and out-of-sample accuracy analysis

*Panel A: In-sample and out-of-sample accuracy of regression models*

| Models | In-sample accuracy | | Out-of-sample accuracy | |
|---|---|---|---|---|
| | Correlation Accuracy | Residual Std. error | Correlation Accuracy | MSFE |
| Univariate | 10% | 1.002 | 5% | 0.97 |
| Univariate with NBER dummy | 12% | 0.91 | 12% | 1.56 |
| Multivariate with risk factors | 76% | 1.66 | 57% | 0.67 |
| Fama-French three-factors | 76%[*] | 0.66[*] | 64%[*] | 0.48[*] |

Note: * indicates best predictive model

*Panel B: In-sample and out-of-sample accuracy of Housing risk beta with Fama-French three factors*

| Dependent variable | In-sample accuracy | | Out-of-sample accuracy | |
|---|---|---|---|---|
| | Correlation Accuracy | Residual Std. error | Correlation Accuracy | MSFE |
| All REIT total return | 74% | 0.66 | 72% | 0.48 |
| All REIT equity return | 71% | 0.69 | 73% | 0.53 |
| Real Estate50 equity returns | 67% | 0.70 | 72% | 0.80 |

error are used as the model selection criteria for in-sample model comparison, the Fama-French model is the preferred model.

The Fama-French model also appears the best performing out-of-sample model according to the MSFE and correlation accuracy criteria. Looking over the out-of-sample portion of panel A, the Fama-French model's directional accuracy is 64%, which is the highest among the models. The model is also the best predictive model considering its lowest value of MSFE. Panel B reports the in-sample and the out-of-sample evaluation of the Fama-French models with the housing risk index. The model exhibits a superior forecasting accuracy in one period ahead prediction for all REIT portfolios, where the average directional accuracy lies between 67% and 74%.

**5.6 Prediction of U.S. economic condition**

After confirming the robustness of housing beta and the best predictive model, we explore the relationship between the housing risk index and measure of U.S. economic and financial conditions. The aim is to examine if the proposed housing risk index can predict the U.S. business cycle fluctuations using six macroeconomic and financial indexes as the dependent variables in OLS regression, in which the model estimates the alternative predictive ability of the proposed housing risk index. There are a significant number of empirical studies investigating the role of U.S. housing sectors and the REIT related variables to measure the U.S. business cycles, as in Christidou and Fountas (2018), Cauley, Pavlov, and Schwartz (2007), Davis and Heathcote (2005), Leamer (2015), and Strobel, Nguyen Thanh, & Lee (2018). Leamer (2015) shows that the housing market accurately predicts eight U.S. economic recessions in the post-war II periods.

The first measure is the National Bureau of Economic Research's (NBER) recession indicators. The indicator is a dummy variable representing periods of expansion for the value of zero, and recession periods take value one. The second one is the Chicago Fed National Activity Index (CFNAI) and the Smoothed Recession Probabilities (SRP) use to measure the overall economic activity and the business cycle fluctuations. The CFNAI is a weighted average of 85 existing monthly indicators of US aggregate economic activity, while SRPs are the estimates of a dynamic-factor Markov-switching model proposed by Chauvet (1998). The SRP index relies on four monthly coincident variables, such as non-farm payroll employment, industrial production index, real personal income excluding transfer payments, and real manufacturing and trade sales.

Finally, the analysis looks beyond the aggregate U.S. economic conditions by considering the regional economic fluctuations. The regional counterpart to the Chicago Fed National Activity Index (CFNAI) is the Midwest Economy Index (MEI), which measures the growth in nonfarm business activity in Illinois, Indiana, Iowa, Michigan, and Wisconsin. The final proxy is the Kansas City Financial Stress Index (KCFSI), which captures the critical aspects of eleven U.S. financial market variables. Developed by Hakkio and Keeton (2009), positive values of KCFSI coincide with periods above the long-run average of financial stress, while negative values indicate the financial stress is below the long-run average.

The predictive analysis estimates the following regression:

$$Index_{t+1} = \alpha + \beta H_t + \varepsilon_{t+1} \tag{15}$$

Where $Index_{t+1}$ denotes one of the six U.S. economic and financial condition measures at time $t+1$ and $H_t$ is the U.S. housing risk index at time $t$. Table 3.8 shows the results of regression equation (15) for each of the six macroeconomic and financial condition indexes. As findings show, housing risk beta is significantly related to future business cycles and financial conditions.

The positive coefficient of housing risk betas predicts official periods of NBER economic recessions and the likelihood for the periods of higher financial stress in the KCFSI. The positive beta also predicts higher U.S. recession probabilities. For the CFNAI and MEI, the negative beta implies the below-average growth, lower nonfarm business activities, and the economic contraction. The negative beta is also related to the periods of lower production in IPG. The predictive power of housing risk beta for all cases remain consistent as predicted. Figure 3.8 shows the reliability of the proposed housing index's prediction with real-time economic and financial predictors. These findings are also consistent with the forecasting ability of aggregate systemic risk factors of Allen, Bali, and Tang (2012), who show that financial sector-specific aggregate systemic risk measures significantly forecast the U.S. and European macroeconomic conditions.

Table 3.8: Regression of U.S. economic condition using housing risk index

====================================================================

| | Dependent variable | | | | | |
|---|---|---|---|---|---|---|
| | NBER (1) | CFNAI (2) | MEI (3) | KCFSI (4) | IPG (5) | US-SRP (6) |
| Housing Risk Index | 0.026*** (0.006) | -0.111*** (0.018) | −0.131*** (0.017) | 0.111*** (0.019) | -0.040*** (0.013) | 2.44*** (0.419) |
| Constant | 0.125*** (0.014) | 0.0001 (0.041) | 0.007 (0.043) | -0.020 (0.052) | 0.186*** (0.030) | 8.586*** 0.947) |
| Observations | 575 | 575 | 511 | 347 | 575 | 575 |
| R2 | 0.031 | 0.062 | 0.095 | 0.089 | 0.016 | 0.056 |

====================================================================

Note:  *p<0.1; **p<0.05; ***p<0.01

$$Index_{t+1} = \alpha + \beta H_t + \varepsilon_{t+1} \qquad (15)$$

Where $Index_{t+1}$ denotes one of the six monthly measures of the US economic and financial index at time $t+1$: The NBER based recession indicators for the United States (NBER), The Chicago Fed National Activity Index (CFNAI) to measure overall US economic activity and related inflationary pressure, The Midwest Economy Index (MEI) to measure growth in nonfarm business activity in the Seventh Federal Reserve District (Illinois, Indiana, Iowa, Michigan, and Wisconsin), The Kansas City Financial Stress Index (KCFSI) to measure the current level of US financial stress, US industrial production growth (IPG), and the Smoothed recession probabilities (SRP) for the United States. $H_t$ is the US housing risk index at time t. The table shows the estimates of the slope and intercept coefficient, $R^2$ statistics, F-statistics, and t-statics. *, **, and *** indicate the statistical significance level at the 10%, 5 % and 1% levels of significance, respectively.

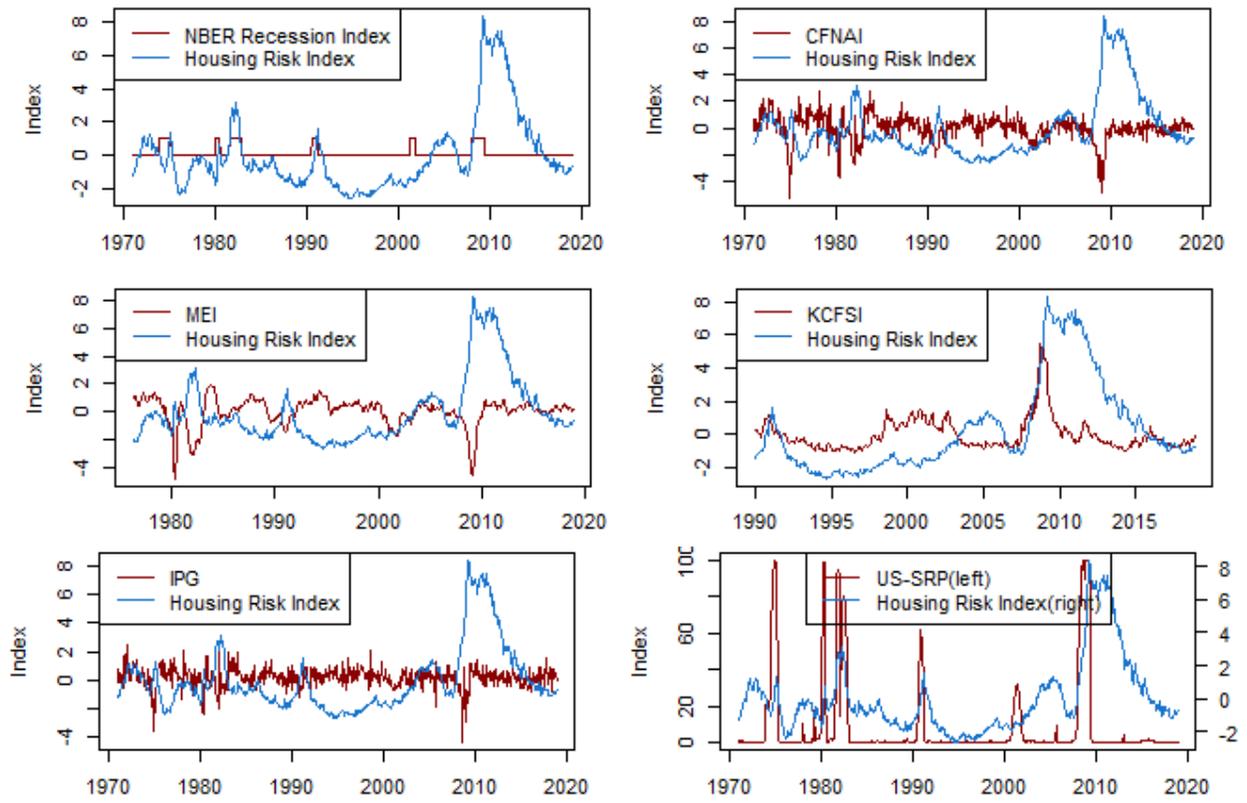

Figure 3.8: Relationship between the housing risk and the U.S. indices

## 6. Conclusion

This chapter estimates Merton's (1973) intertemporal capital asset pricing model and the Fama-French three-factor model using a proposed housing risk factor. The housing risk index is the first principal component of the ten time-varying conditional volatilities, which serves as an alternative aggregate measure of the U.S. housing sector volatilities. A higher value of the U.S. housing risk index is a significant leading indicator of downturns in U.S. business cycles, leaves investors in higher-risk positions, which requires a higher risk premium in their investment decisions. Due to the vital role of the housing sectors in the U.S. economy, REIT investors are likely to focus more on the risk associated with the aggregate macroeconomic condition and the various components of housing sectors, which potentially provide a guide to their optimal investment decisions.

The findings show a significant positive relation between housing risk index and REIT equity returns across all regression specifications. The estimated coefficient of housing risk beta is economically meaningful in magnitude, where values lie between 0.019 and 0.174 depending on the specification of regression models and the sample periods. The sub-sample and out-of-sample estimation of betas validate the theoretical consistency of Fama-French and the ICAPM models. Although the aggregate index of housing sector volatilities positively priced, it only explains a small fraction of the time-series variation of REIT return in univariate cases. After controlling for the standard macroeconomic-finance variables from the asset pricing literature, multivariate OLS and quantile regression models successfully capture a larger proportion of the variation of REIT returns. The quantile estimates of all Fama-French risk factors are also positive and statistically significant across the distributions, while the relationship is positive and significant only on the REIT return distributions that lie in the upper quantile.

# 8. Appendix

## I. Data construction and sources

| Data | Time-period | Source |
|------|-------------|--------|
| HOUST; PERMIT. | | |
| HPRICE; MORINT; | 1971-2018 | Federal Reserve Bank of St. Louis |
| HSTMW; HSTNE. | | |
| HSTSOU; HSTW | | |
| HSOLD; | 1971-2018 | U.S. census bureau's historical data |
| SATSOLD | | https://www.census.gov/construction/nrs/historical_data/index.html |
| SMB: Small Minus Big. HML: High Minus Low. MOM: Monthly Momentum Factor. CrdSpr: Credit spread. D12: 12-month moving sums of dividends paid on the S&P 500 index: E12: 12-month moving sums of earnings; b/m: Book-to-Market Ratio. lty: long-term government bond yield. ntis: Net Equity Expansion. svar: Stock Variance. dfy: Default Yield Spread. | 1971-2018 | Goyal's website (Welch and Goyal, 2008) http://www.hec.unil.ch/agoyal/ |
| Skew: Average skewness of U.S. firm level stock returns | 1971-2018 | Jondeau, Zhang, and Zhu (2019) |
| Macroeconomic data set McCracken and Ng (2020) | 1971-2018 | FED St Louis |

## II. List of macroeconomic variables

**Group 1: National Income and Product Accounts (NIPA)**

1. Real Gross Domestic Product, 3 Decimal (Billions of Chained 2012 Dollars)
2. Consumption Real Personal Consumption Expenditures (Billions of Chained 2012 Dollars)
3. Real personal consumption expenditures: Durable goods (Billions of Chained 2012 Dollars)
4. Real Personal Consumption Expenditures: Services (Billions of 2012 Dollars),
5. Real Personal Consumption Expenditures: Nondurable Goods (Billions of 2012 Dollars)
6. Investment Real Gross Private Domestic Investment, (Billions of Chained 2012Dollars)
7. Real private fixed investment (Billions of Chained 2012 Dollars)

8. Real Gross Private Domestic Investment: Fixed Investment: Nonresidential: Equipment (Billions of Chained 2012 Dollars)
9. Real private fixed investment: Nonresidential (Billions of Chained 2012 Dollars)
10. Real private fixed investment: Residential (Billions of Chained 2012 Dollars)
11. Shares of gross domestic product: Gross private domestic investment: Change in private inventories (Percent)
12. Real Government Consumption Expenditures & Gross Investment (Billions of Chained 2012 Dollars)
13. Real Government Consumption Expenditures and Gross Investment: Federal (Percent Change from Preceding Period)
14. Real Gov Receipts Real Federal Government Current Receipts (Billions of Chained 2012 Dollars)
15. Real government state and local consumption expenditures (Billions of Chained 2012 Dollars)
16. Real Exports of Goods & Services (Billions of Chained 2012 Dollars)
17. Real Imports of Goods & Services (Billions of Chained 2012 Dollars)
18. Real Disposable Personal Income (Billions of Chained 2012 Dollars)
19. Nonfarm Business Sector: Real Output (Index 2012=100)
20. Business Sector: Real Output (Index 2012=100)
21. Manufacturing Sector: Real Output (Index 2012=100)

## Group 2: Industrial Production

1. Industrial Production Index (Index 2012=100)
2. Industrial Production: Final Products (Market Group) (Index 2012=100)
3. Industrial Production: Consumer Goods (Index 2012=100)
4. Industrial Production: Materials (Index 2012=100)
5. Industrial Production: Durable Materials (Index 2012=100)
6. Industrial Production: Nondurable Materials (Index 2012=100)
7. Industrial Production: Durable Consumer Goods (Index 2012=100)
8. Industrial Production: Durable Goods: Automotive products (Index 2012=100)
9. Industrial Production: Nondurable Consumer Goods (Index 2012=100)
10. Industrial Production: Business Equipment (Index 2012=100)
11. Industrial Production: Consumer energy products (Index 2012=100)
12. Capacity Utilization: Total Industry (Percent of Capacity)
13. Capacity Utilization: Manufacturing (SIC) (Percent of Capacity)
14. Industrial Production: Manufacturing (SIC) (Index 2012=100)
15. Industrial Production: Residential Utilities (Index 2012=100)
16. Industrial Production: Fuels (Index 2012=100)

## Group 3: Employment and Unemployment

1. All Employees: Total nonfarm (Thousands of Persons)
2. All Employees: Total Private Industries (Thousands of Persons)
3. All Employees: Manufacturing (Thousands of Persons)
4. All Employees: Service-Providing Industries (Thousands of Persons)
5. All Employees: Goods-Producing Industries (Thousands of Persons)
6. All Employees: Durable goods (Thousands of Persons)
7. All Employees: Nondurable goods (Thousands of Persons)
8. All Employees: Construction (Thousands of Persons)
9. All Employees: Education & Health Services (Thousands of Persons)
10. All Employees: Financial Activities (Thousands of Persons)
11. All Employees: Information Services (Thousands of Persons)
12. All Employees: Professional & Business Services (Thousands of Persons)
13. Civilian Employment (Thousands of Persons)
14. Civilian Labor Force Participation Rate (Percent)
15. Civilian Unemployment Rate (Percent)
16. Business Sector: Hours of All Persons (Index 2012=100)
17. Manufacturing Sector: Hours of All Persons (Index 2012=100)
18. Nonfarm Business Sector: Hours of All Persons (Index 2012=100)
19. Average Weekly Hours of Production and Nonsupervisory Employees: Manufacturing (Hours)
20. Weekly Hours of Production And Nonsupervisory Employees: Total private (Hours)
21. Help-Wanted Index

**Group 4: Prices**

1. Gross Domestic Product: Chain-type Price Index (Index 2012=100)
2. Gross Private Domestic Investment: Chain-type Price Index (Index 2012=100)
3. Consumer Price Index for All Urban Consumers: All Items (Index 1982-84=100)
4. Consumer Price Index for All Urban Consumers: All Items Less Food & Energy (Index 1982-84=100)
5. Producer Price Index for All Commodities (Index 1982=100)

**Group 5: Interest Rates**

1. Effective Federal Funds Rate (Percent)
2. 10-Year Treasury Constant Maturity Rate (Percent)
3. 30-Year Conventional Mortgage Rate© (Percent)
4. Moody's Seasoned Aaa Corporate Bond Yield© (Percent)
5. Moody's Seasoned Baa Corporate Bond Yield© (Percent)

**Group 6: Money and Credit**

1. St. Louis Adjusted Monetary Base (Billions of 1982-84 Dollars)
2. Real Institutional Money Funds (Billions of 2012 Dollars)
3. Real M1 Money Stock (Billions of 1982-84 Dollars)
4. Real M2 Money Stock (Billions of 1982-84 Dollars)
5. Total Reserves of Depository Institutions (Billions of Dollars)
6. Reserves of Depository Institutions, Nonborrowed (Millions of Dollars)

**Group 7: Stock Markets**

1. S&P's Common Stock Price Index: Composite
2. S&P's Common Stock Price Index: Industrials